\documentclass[aps,prb,twocolumn,groupedaddress,amsmath,amssymb,floatfix,showpacs,citeautoscript,longbibliography]{revtex4-2}
\usepackage[T1]{fontenc}
\usepackage{graphicx}
\usepackage{float}
\usepackage{bm}
\usepackage{braket}
\usepackage{breqn}
\usepackage[dvipsnames]{xcolor}

\begin{document}
\title{Spin Qubits in Photon-Coupled Microwave Cavities}
\author{Samuel Johnson}
\author{Nancy Sandler}
\email{sandler@ohio.edu}
\affiliation{Department of Physics and Astronomy and Nanoscale and Quantum Phenomena Institute -- Ohio University -- Athens -- Ohio 45701 -- USA}

\begin{abstract}
Electron spin qubits in microwave cavities provide a promising platform for scalable quantum computing hardware, leveraging long coherence times, charge-noise robustness and cavity mediated qubit-qubit interactions. While the strong spin-photon coupling regime is accessible via on-chip micromagnets, scaling conventional architectures by placing multiple qubits within a single shared resonator degrades transmission amplitudes, hence limiting large-scale efficiency. To overcome this limitation, we analyze
a modular architecture where individual cavities containing a limited number of qubits are coupled via single-photon-exchange waveguides. Using input/output theory, we compute the transmission amplitudes for networks of two and three coupled cavities in various configurations. We map out the distinct physical regimes accessible by tuning key system parameters, offering a viable pathway for scalable cavity-based quantum spin qubit networks.
\end{abstract}

\maketitle

\section{Introduction}
\label{sec1}
The realization of a scalable quantum computer requires qubits that can be coupled, entangled, and operated through a universal set of logic gates~\cite{DiVincenzo2000}.  Among the many physical implementations currently under investigation, electron spins in Si/SiGe heterostructures stand out as a particularly attractive candidate for integration with existing semiconductor manufacturing facilities.
These systems benefit from strong robustness against charge noise, with spin qubits in Silicon exhibiting coherence times on the order of seconds, far exceeding those of most competing semiconductor-based platforms~\cite{Yang_2013,Burkard2023}. Moreover, high-fidelity single and two-qubit gate operations in these structures, as well as spin-state readout, have also been demonstrated ~\cite{Yoneda_2017,Xue_2022,Noiri_2022,Huang_2024,Takeda_2024,Yang_2019,Wu2024-hj}. 

Extending the range of spin-spin interactions remains a central challenge for scalable spin-qubit architectures. Cavity quantum electrodynamics (cQED) offers an established and versatile route to this problem: qubits are coherently coupled to the electromagnetic field of a microwave resonator, allowing spatially separated spins to interact via virtual or real cavity photons over macroscopic distances. More recently, spin shuttling has emerged as a complementary strategy, in which electrons are displaced using electric gate potentials while maintaining high spin fidelity~\cite{DeSmet2025}; this approach, however, requires preserving coherence throughout the transport process, over the full length of the desired interaction range~\cite{Bosco2024,Foster2025,Pazhedath2025}.

In the cQED approach, electrons are confined to semiconductor quantum dots housed within microwave cavities~\cite{Petersson_2012,Beaudoin_2016,Viennot_2015}. A particularly successful realization~\cite{Borjans_2019} employs micromagnets to generate local magnetic field gradients that enable spin-photon coupling, with cavity photons subsequently mediating interactions between spatially separated qubits~\cite{Mi_2017,Benito_2017,Mielke2021,Mills_2022,Mi_2018,Samkharadze_2018}.
This platform -that in principle, supports all-to-all connectivity among multiple qubits-; has reached the strong-coupling regime and multi-qubit interactions have already been demonstrated for small qubit numbers~\cite{Borjans_2019,Dijkema2025,Philips2022, Thorvaldson_2025}. Scaling this success to larger systems, however, remains an open problem.

The conventional approach places multiple qubits in a single shared cavity, where interactions are mediated by a common cavity field mode. As we show below, this configuration suffers a rapid reduction in output transmission amplitude as the qubit number grows, limiting its viability for large-scale architectures. Here, we investigate an alternative modular scheme in which a small number of qubits are housed in separated resonator cavities, with scalability achieved by connecting multiple cavities via capacitive couplers~\cite{Schmidt_2013, Kollar_2019,PhysRevA.86.023837}, in the spirit of related theoretical and experimental work~\cite{Wang2013, Ogden_2008, Marcos_2012, Greentree_2006, Angelakis_2007, Hartmann_2006}. This design offers improved individual qubit addressability and avoids the signal degradation inherent to large shared-cavity arrays.

We analyze the first steps of this scaling approach by extending the single-cavity model of Ref.~\cite{Benito_2017} to systems of two and three coupled cavities in various configurations. We employ a fully quantum-mechanical treatment of the cavity field together with input-output theory to obtain experimentally verifiable signatures of the system's behavior. 

The paper is organized as follows. 
Sec.~\ref{sec2} reviews the single-qubit, single-cavity model of Ref.~\cite{Benito_2017}, reproducing several of its results as a starting point to examine how transmission evolves as additional qubits are introduced. Sec.~\ref{sec3} presents results for two and three coupled cavities, each containing a single qubit, across various coupling configurations and parameter regimes. Sec.~\ref{sec4} considers two coupled cavities with differing numbers of qubits. We close with our conclusions in Sec.~\ref{sec5}.

\section{Single Cavity Model}
\label{sec2}

 The qubit is realized by a double-quantum-dot (DQD) structure whose low-energy physics is captured by a two-dimensional Hilbert space spanned by the two charge configurations corresponding to the electron occupying either dot. The Hamiltonian of this charge qubit is governed by two experimentally tunable parameters: the interdot detuning $\epsilon$, defined as the on-site energy difference between the two charge states, and the tunnel coupling $t_c$, which controls the hybridization between them. Both quantities can be independently controlled via external voltage gates, allowing precise in-situ tuning of the qubit's energy splitting and degree of charge admixture.

The corresponding Hamiltonian takes the form:
\begin{equation}
    H_0 = \mathbb{I}_\sigma\otimes\frac{1}{2}(\epsilon\tau_z + 2t_c\tau_x)
    \label{eq:H0}
\end{equation}
\noindent where $\tau_\alpha$ represent Pauli operators acting in position space, and $\mathbb{I}_\sigma$ represents the identity operator in spin space.

The introduction of a static magnetic field -realized experimentally via on-chip micromagnets- plays a crucial role in this architecture. The field gradient across the DQD generates a position-dependent Zeeman splitting so that the charge degrees of freedom become entangled with the spin. As a consequence, the charge-photon coupling translates directly into an effective spin-photon coupling, enabling coherent photon-mediated spin manipulation. In addition, a homogeneous component $B_z$ along the line that joins the dots provides control on the resonance conditions.

The Hamiltonian, including these fields, is
\begin{equation}
    H_b = \frac{1}{2}(\mathbb{I}_\sigma\otimes\epsilon\tau_z +  \mathbb{I}_\sigma\otimes 2t_c\tau_x + B_z\sigma_z\otimes\mathbb{I}_\tau + B_x\sigma_x\tau_z).
    \label{eq:Hb}
\end{equation}
where the $\sigma_\alpha$ operators represent the Pauli operators acting on spin space, $\mathbb{I}_\tau$ is the identity in position space and the magnetic fields are in energy units. 
\begin{figure}
     \centering
    \includegraphics[width=0.48\textwidth]{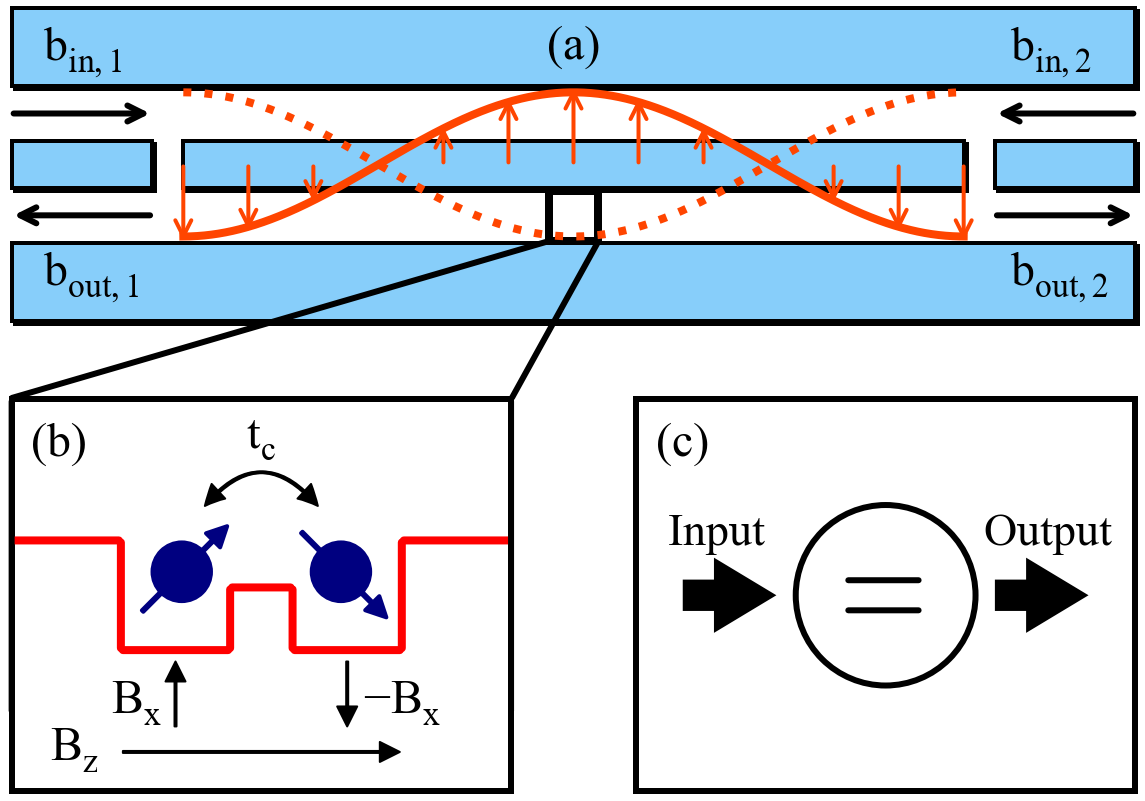}
     \hfill
        \caption{(a) Rough sketch of a coplanar waveguide resonator with a single DQD qubit, shown magnified in (b), placed inside. The electric field associated with a cavity photon is shown as a standing wave in red. The resonator is capacitively coupled to external ports on both sides that allow photons to be exchanged with the outside; input and output fields are labeled as $b_{in}$, $b_{out}$. (b) Schematic diagram of a DQD qubit. A single electron is placed inside a double quantum dot, represented in red as a double potential well. The magnetic moment of the electron will align with the homogeneous magnetic field $B_z$ and the inhomogeneous field $B_x$. (c) A schematic representation of the combined qubit-cavity system. A signal is sent in at the input port, and a transmission amplitude is calculated by measuring the output on the other side. The input port is described by the fields $b_{in,1}$, $b_{out,1}$ while the fields $b_{in,2}$, $b_{out,2}$ describe the output port.}
        \label{fig:single-cavity-schematic}
\end{figure}
The DQD is placed inside a single-mode microwave resonator, whose field is described by the Hamiltonian  $H_{ph} = \omega_ca^\dagger a$ (in units of $\hbar =1$), where $a^\dagger$ and $a$ are the bosonic creation and annihilation operators acting on the coherent state space of the cavity photons, and $\omega_c$ is the bare resonator frequency. Coupling between the DQD charge degree of freedom and the cavity electric field arises from the electric dipole interaction. In the charge basis, this takes the form:
\begin{equation}
   H_c = g_c(a + a^\dagger)\mathbb{I}_\sigma\otimes\tau_z,
   \label{eq:Hc}
\end{equation}

\noindent where $g_c$ is the bare charge-photon coupling strength. Using the eigenbasis of $H_b$ (see Appendix A for a derivation of eigenstates for $\epsilon = 0$), this expression becomes
\begin{equation}
    H_c = g_c(a + a^\dagger) \sum_{n,m=0}^{3} d_{nm}\ket{n}\bra{m}.
    \label{eq:Hcb}
\end{equation}
\noindent where $(n,m)$ label the various energy eigenstates and $d_{n,m}$ are the dipole matrix elements. The full single-qubit Hamiltonian is then
\begin{equation}
    H = H_b + H_c + H_{ph}.
\label{eq:H}
\end{equation}

To account for the inevitable coupling of the system to its environment, dissipative terms describing photon loss from the cavity at the rate $\kappa$, and qubit relaxation at rate $\gamma$ are added to Eq.~(\ref{eq:H}). The dynamics of the system are thus governed by the Lindblad master equation. Experimentally accessible information about the system is obtained through cavity transmission, which we compute via input-output theory that considers the qubit-cavity as an open system connected to external photon baths~\cite{Gardiner1985}. Here, the bath is treated dynamically, and input and output fields ($b_{in}$, $b_{out}$) contain the information about the system's internal state via the boundary condition $b_{out} - b_{in} = \sqrt{\kappa} a$. By applying the quantum Langevin equation to the relevant operators for the single cavity Hamiltonian (Eq.~(\ref{eq:H})), we derive the equations of motion for the photon operators $a$ and $a^{\dagger}$ as well as the qubit operators $\sigma_{nm} = \ket{n}\bra{m}$:
\begin{dmath}
    \dot{a} = -ig_c\sum_{n,m=0}^3d_{nm}\sigma_{nm} - i\omega_ca - \sqrt{\kappa_1}b_{in,1}(t) 
    -\sqrt{\kappa_2}b_{in,2}(t) + \frac{\kappa}{2}a 
  \label{eq:adot}
\end{dmath}
\begin{dmath}
    \dot{\sigma}_{nm} = -i(E_m - E_n)\sigma_{nm} - \sum_{n',m' = 0}^3\gamma_{nm,n'm'}\sigma_{n'm'} 
    - ig_c(a + a^{\dagger})d_{nm}(P_n - P_m)
    \label{eq:sigmadot}
\end{dmath}
Here, $E_n$ represents the energy of the $n^{th}$ eigenstate of $H_b$, and $P_n = \langle \sigma_{nn}\rangle$ is the average population of the $n-th$ energy level. The decoherence superoperator $\gamma$ accounts for dephasing due to charge noise ~\cite{Benito_2017}. The operators $b_{in,1}$, $b_{in,2}$ stand for input fields applied at the external ports (labeled 1 and 2) on the two sides of the cavity. The parameter $\kappa_i$ represents the strength of the coupling between the external electromagnetic bath and the cavity through each individual port. We consider that the overall damping $\kappa$ occurs at these ports only, resulting into $\kappa/2 = \kappa_1 = \kappa_2$.

The system is driven by a monochromatic field at frequency $\omega_R$. Using the rotating wave approximation -valid near the resonant regime-, we obtain the time-averaged values for the system operators: 
\begin{dmath}
    \Bar{\dot{a}} = -ig_c\sum_{n=0}^2\sum_{j = 1}^{3-n}d_{n,n+j}\Bar{\sigma}_{n,n+j} + i\Delta \Bar{a} - \sqrt{\kappa_1}\Bar{b}_{in,1} - \sqrt{\kappa_2}\Bar{b}_{in,2} - \frac{\kappa}{2}\Bar{a}
    \label{eq:abardot}
\end{dmath}
\begin{dmath}
    \Bar{\dot{\sigma}}_{n,n+j} = -\sum_{n',j'}\gamma_{n,n+j,n',n'+j'}\Bar{\sigma}_{n',n'+j'} -i(E_{n+j} - E_n - \omega_R)\Bar{\sigma}_{n,n+j} - ig_c\Bar{a}d_{n,n+j}(P_n - P_{n+j})
    \label{eq:sigmabardot}
\end{dmath}
\noindent where we introduced the parameter $\Delta = \omega_R - \omega_c$ to represent deviations from resonance. A similar set of equations is obtained in terms of the $\Bar{b}_{out,i}$ fields~\cite{Gardiner1985}.

These equations can be simplified further by considering transitions between two isolated states in the regime where $\omega_c \approx B_Z$. We label these transitions as $0\leftrightarrow1$ and $0\leftrightarrow2$ transitions, reducing the set of equations to those for the operators $a$, $\sigma_{01}$, $\sigma_{02}$.

In the stationary limit, the relationship between incoming and outgoing fields $\Bar{b}_{out,i} - \Bar{b}_{in,i} = \sqrt{\kappa_i}\Bar{a}$ can be used to calculate the transmission amplitude through the cavity as $A_1 = \bar{b}_{out,2}/\bar{b}_{in,1}$.
\begin{equation}
    A_1 = \frac{-i \sqrt{\kappa_1\kappa_2}}{\Delta  -g_c(d_{01}\chi_{01} + d_{02}\chi_{02})+\frac{i\kappa}{2}}
    \label{eq:A1}
\end{equation}
\begin{figure}
    \centering
    \includegraphics[width=0.48\textwidth]{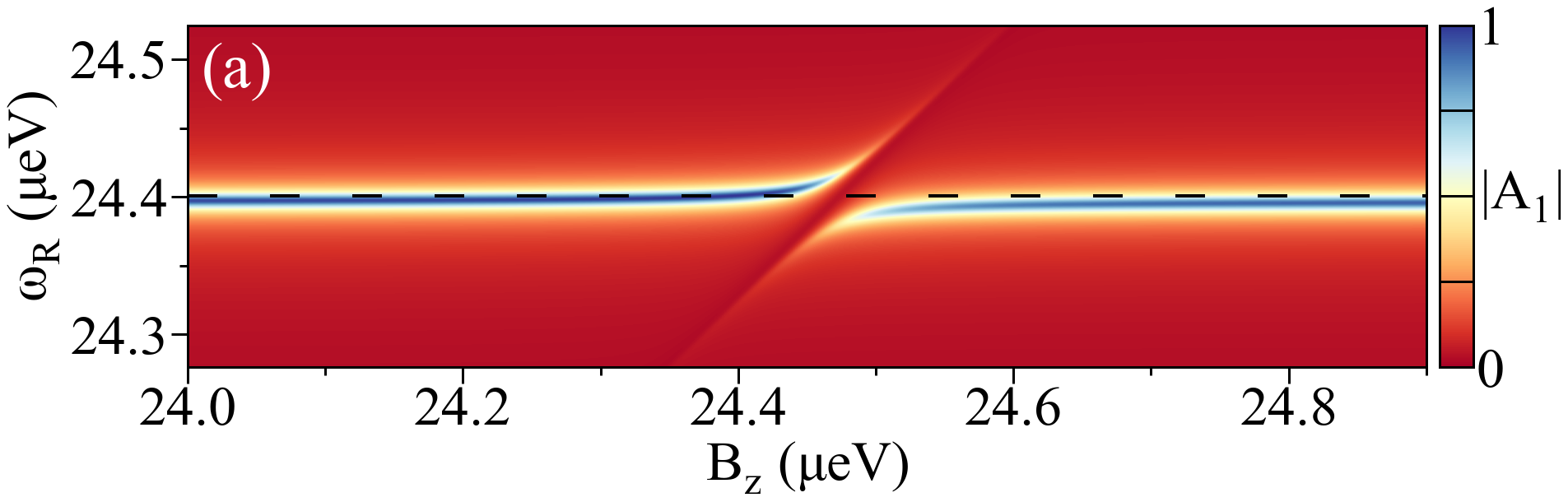}
    \hfill
    \includegraphics[width=0.48\textwidth]{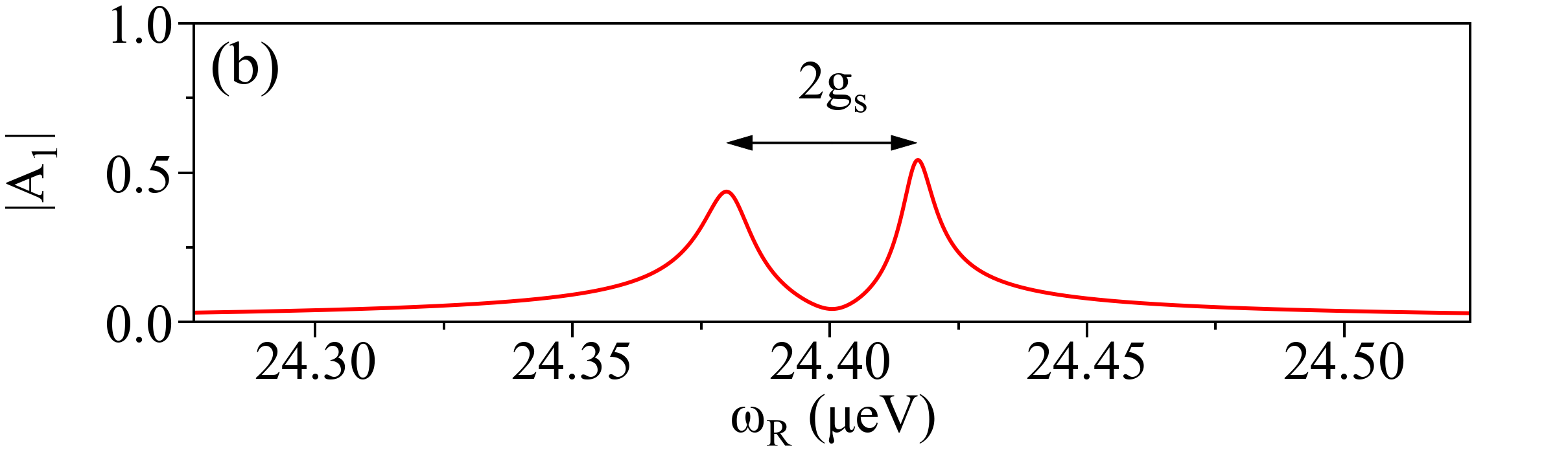}
        \caption{(a) Transmission amplitude through a single cavity system as a function of $B_z$ and $\omega_R$. The value of $\omega_c$ is plotted as a dashed line. (b) Transmission amplitude vs driving frequency at the resonant magnetic field value $B_z^{res} \approx$ 24.47 $\mu$eV. In both plots $\epsilon$ = 0, $B_x$ = 1.6 $\mu$eV, $t_c$ = 16 $\mu$eV, $\gamma_c/2\pi$ = 90 MHz, $g_c/2\pi$ = 50 MHz, $\kappa_1/2\pi$ = $\kappa_2/2\pi$ = 0.9 MHz, $\omega_c/2\pi$ = 5.9 GHz.}
        \label{fig:trans_1cavity1dot}
\end{figure}
 \noindent In this expression, the susceptibilities $\chi$ are defined as $\bar{\sigma}_{nm} = \chi_{nm}\bar{a}$. Note that the presence of the qubit renormalizes the cavity frequency as expected for a loaded cavity~\cite{EMJackson}.
 
 By appropriate tuning of the magnetic field $B_z$, the DQD qubit is brought into resonance with the cavity field, and a corresponding Rabi oscillation ensues ~\cite{JC1963,Brune1996}

Fig.~\ref{fig:trans_1cavity1dot} shows results for the absolute value of the transmission amplitude $|A_1|$ through a cavity containing a single qubit, depicting a 
 Rabi splitting in the transmission spectrum when the qubit is in resonance with the cavity photon.
 
 The spin-cavity coupling rate $g_s$ is obtained by measuring the magnitude of the Rabi splitting ~\cite{Benito_2017}, represented in  Fig.~\ref{fig:trans_1cavity1dot}(b) by the distance between the peaks of the transmission amplitude at the resonant magnetic field value $B_z^{res}$. The resonant value $B_z^{res}$ is determined by the condition that the difference between two qubit energy levels should match the photon energy $E_{1(2)} - E_0 = \omega_c$, resulting in the expression:
\begin{equation}
    B_z^{res} = \omega_c\sqrt{1 - \frac{B_x^2}{\omega_c - 4t_c^2}}.
    \label{eq:Bzres}
\end{equation}

A standard scaling approach for this configuration consists of adding multiple DQDs in the same cavity. Within the method outlined above and assuming no direct qubit-qubit interactions, the resulting transmission amplitude for a single cavity with $N$ qubits is given by:
\begin{equation}
    A_{1}^{(N)} = \frac{-i\sqrt{\kappa_1\kappa_2}}{\Delta  - \sum_{j=1}^Ng_{c,j}(d_{01,j}\chi_{01,j} + d_{02,j}\chi_{02,j})+i\frac{\kappa}{2}}
    \label{eq:An}
\end{equation}

\begin{figure}
    \centering
    \includegraphics[width=0.48\textwidth]{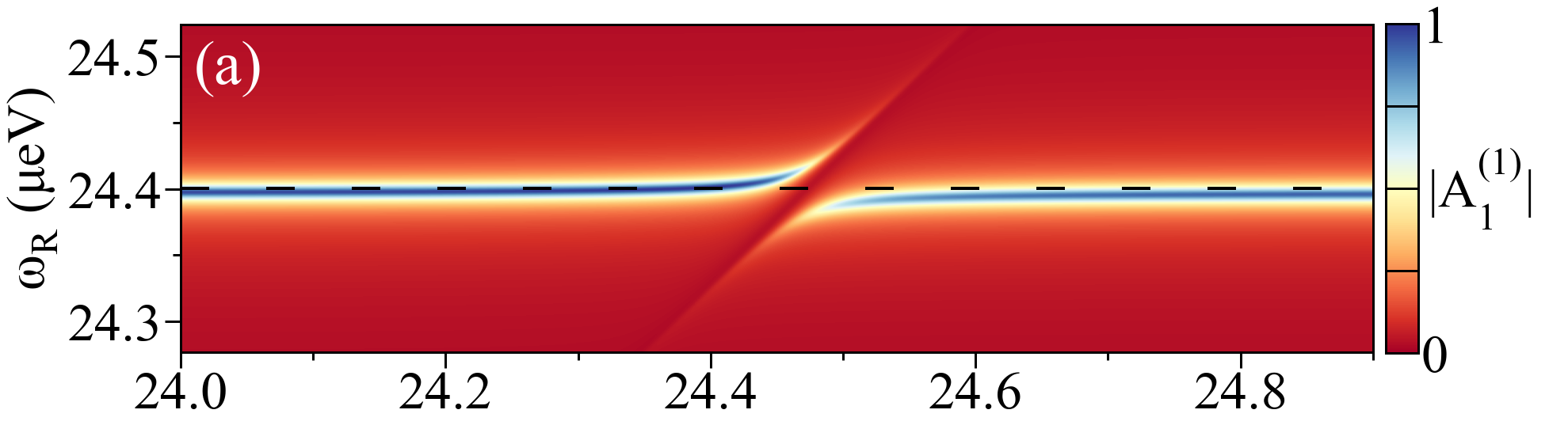}
    \hfill
    \includegraphics[width=0.48\textwidth]{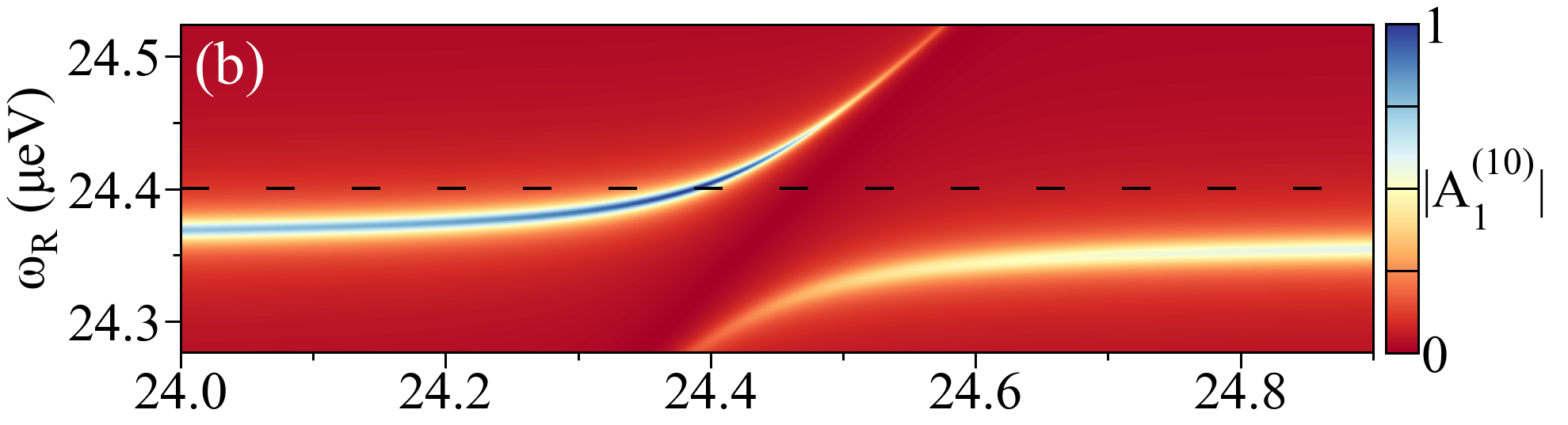}
    \hfill
    \includegraphics[width=0.48\textwidth]{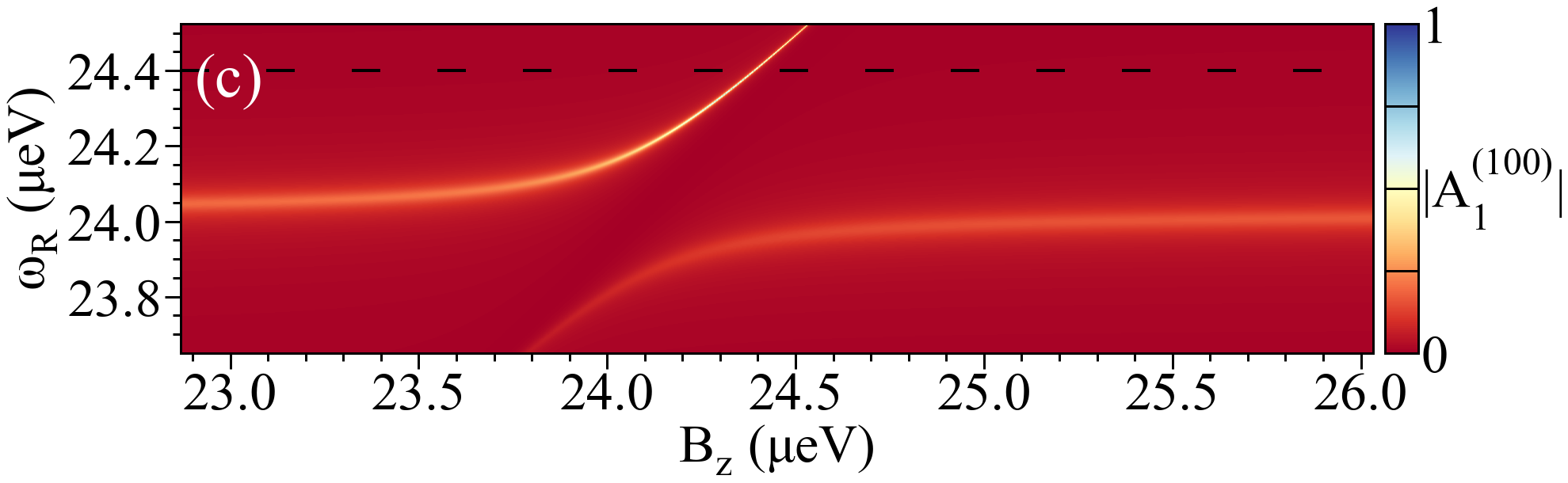}
        \caption{Transmission amplitude through a single cavity for (a) 1 qubit, (b) 10 qubit, and (c) 100 qubit systems. Note the change in scale in subplot (c). In in all plots $\epsilon$ = 0, $B_x$ = 1.6 $\mu$eV, $t_c$ = 16 $\mu$eV, $\gamma_c/2\pi$ = 90 MHz, $g_c/2\pi$ = 50 MHz, $\kappa_1/2\pi$ = $\kappa_2/2\pi$ = 0.9 MHz., $\omega_c/2\pi$ = 5.9 GHz. All qubits are taken to be identical. The resonant frequency of the empty cavity is marked on the y-axis as a black dashed line.}
        \label{fig:trans_1cavityNdot}
\end{figure}

\noindent Here, the sum over $j$ represents a sum over the $N$ qubits in the resonator cavity. Fig.~\ref{fig:trans_1cavityNdot} shows the transmission through a single cavity containing 1, 10, and 100 qubits with the same qubit parameters. As the number of qubits increases, the overall transmission through the system decreases. This can be seen by comparing Fig.~\ref{fig:trans_1cavityNdot}(a) and Fig.~\ref{fig:trans_1cavityNdot}(c) where the jump from 1 to 100 qubits causes the transmission to weaken by $\sim$80\%.

The addition of new qubits also causes a downward shift in the effective resonant cavity frequency, seen in Fig.~\ref{fig:trans_1cavityNdot} as a gradual decrease in the driving frequency necessary for high transmission. This is accompanied by a shift in the value for the resonant magnetic field required for optimal photon-qubit coupling. In a system with $N$ qubits, the Rabi splitting will now occur at a value of $B_z^{res}$ calculated from Eq.~[\ref{eq:Bzres}] where $\omega_c$ is replaced by $\omega_{c,eff}$ which can be estimated by evaluating the following expression away from resonance:
\begin{equation}
\omega_{c,eff} \approx w_c + Re\left(\sum_{j=1}^Ng_{c,j}(d_{01,j}\chi_{01,j} + d_{02,j}\chi_{02,j})\right)
\end{equation}

Thus, the qubits act as a dispersive load for the cavity, renormalizing its total quality factor $Q$. This has a direct practical implication: the cavity design must be appropriately adjusted to accommodate the modified $Q$ as the number of hosted qubits changes~\cite{EMJackson}. Nevertheless, using the corrected resonance condition, the Rabi splitting can be accurately determined for systems with an arbitrary qubit number $N$. 

As $N$ increases, the collective spin-photon coupling rate $g_s$ grows as $\sqrt{N}$, consistent with the enhancement predicted by the Tavis-Cummings model for $N$ two-level systems interacting coherently with a single cavity mode~\cite{PhysRev.170.379}. It is worth emphasizing, however, that this cooperative enhancement does not translate into improved transmission performance. As shown in Fig.~\ref{fig:trans_1cavityNdot}, the output signal degrades with increasing $N$, thus motivating the modular multi-cavity architecture investigated in the remainder of this work.

\section{Multiple Cavities}
\label{sec3}

As described in the introduction, an alternative scaling scheme for a multi-qubit system can be created by connecting multiple resonator cavities, each containing one or more qubits.
 
The intercavity connections can be achieved via capacitive couplers, as described in Refs.~\cite{Kollar_2019,Painter2023}, which allow photons to move between different cavities. We first focus on the simpler two-cavity system with a single qubit in each cavity and, in Sec.~\ref{sec3B}, we extend the model to three cavities, allowing for different connection schemes.

\subsection{The two-cavity system}
\label{sec3A}

In analogy to the DQDs, we introduce a photonic degree of freedom for each cavity, 
identified by the labels left (L) and right (R) to indicate the cavities' relative positions. 

The Hamiltonian that describes the exchange of photons between the two cavities is:
\begin{equation}
    H_{e,2} = t(a_L^\dagger a_R + a_R^\dagger a_L)
    \label{eq:He}
\end{equation}
\noindent where the parameter $t$ defines the energy associated with the photon transfer. The Hamiltonian governing the dynamics of the two-cavity system is then given by:
\begin{equation}
    H_{sys,2} = H_L + H_R + H_{e,2}
    \label{eq:Hsys}
\end{equation}
where $H_{L,R}$ match the single cavity Hamiltonian given in Eq.~(\ref{eq:H}) with operators labeled by $L,R$ for the left and right cavities respectively.

\begin{figure}[h]
     \centering
     \includegraphics[width=0.48\textwidth]{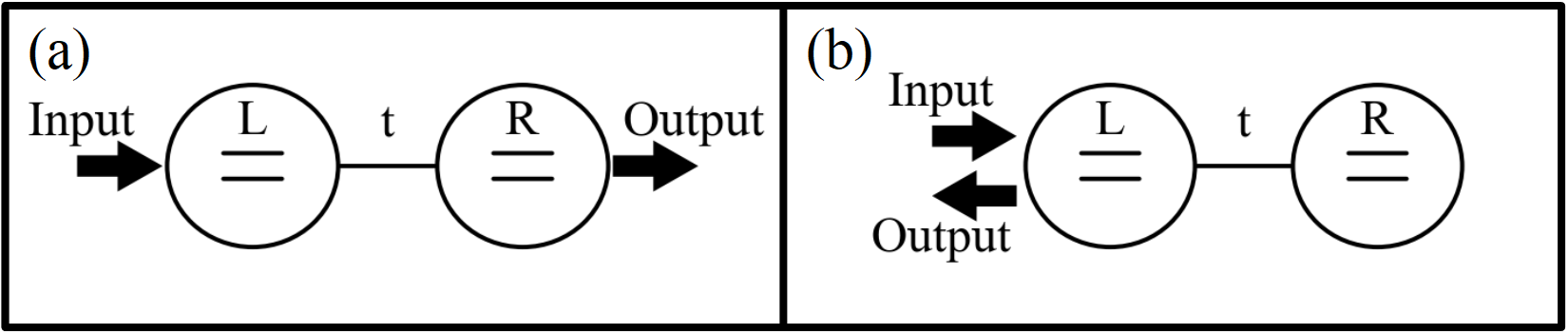}
     \hfill
        \caption{Schematic diagram showing two possible layouts for the two-cavity system. Two microwave resonator cavities (labeled L,R), each containing a single DQD qubit, are connected via a coupling capacitor with coupling strength $t$. An input signal is sent in at the left cavity and the output signal then is measured at either the right cavity (a) or the left cavity (b). These two setups are referred to as type-A and type-B respectively.}
        \label{fig:2cavity_schematic}
\end{figure}

Because the system now contains multiple cavities, there is a choice for how to connect the external ports. The input and output ports can be connected to the left and right cavities separately, the type-A layout, or they can both be connected to a single cavity, the type-B layout. These two systems are shown schematically in Fig.~\ref{fig:2cavity_schematic}. Notice that in the type-B layout, photon dissipation processes described by $\kappa$ are included for only one cavity, as they are associated with the connection to the external ports. 

Following the procedure outlined for the single cavity case, we obtain new equations of motion for the two-cavity system and calculate a transmission amplitude as ${A_2} = \bar{b}_{out,2}/\bar{b}_{in,1}$. For a type-A system the transmission amplitude reads:
\begin{equation}
    A_2^{(A)} = \frac{-i t \sqrt{\kappa_1\kappa_2}}{(\xi_L +i\frac{\kappa_1}{2})(\xi_R + i \frac{\kappa_2}{2}) - t^2}
    \label{eq:A2}
\end{equation}
\noindent where we introduced the variables $\xi_i$ as:
\begin{equation}
    \xi_i = \Delta_i -g_{c,i}(d_{01,i}\chi_{01,i} + d_{02,i}\chi_{02,i}) 
    \label{eq:xii}
\end{equation}
\noindent with $\Delta_i = \omega_R - \omega_{c,i}$ and susceptibilities $\chi_{nm,i}$ for each cavity. 

\begin{figure}
    \centering
    \includegraphics[width=0.48\textwidth]{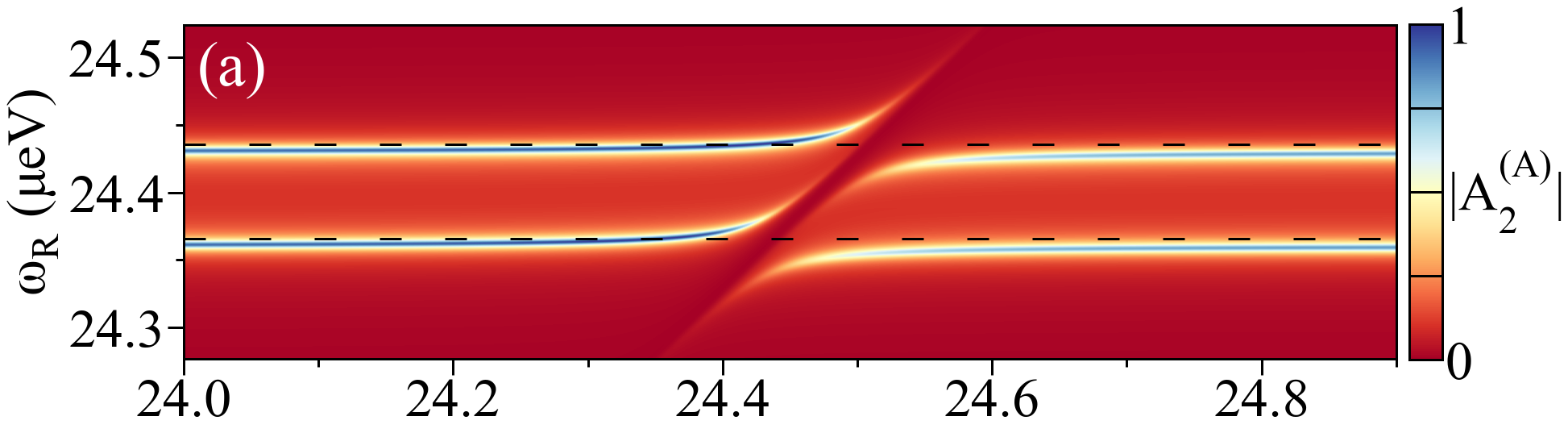}
    \hfill
    \includegraphics[width=0.48\textwidth]{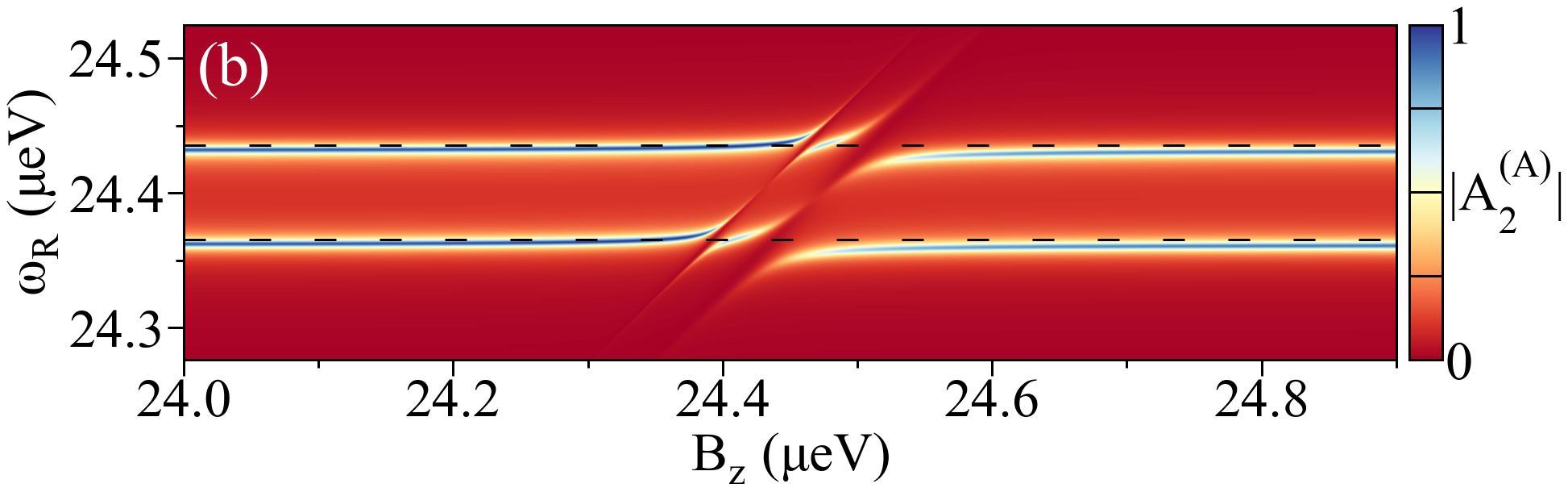}
        \caption{(a) Transmission amplitude through a type-A two-cavity, two-qubit system as a function of $B_z$ and $\omega_R$. In this plot the two qubits have identical parameters with $t_{c,L}$ = $t_{c,R}$ = 16 $\mu$eV, $B_{x,L}$ = $B_{x,R}$ = 1.6 $\mu$eV. In all plots the cavities are identical with $\omega_{c,L}/2\pi$ = $\omega_{c,R}/2\pi$ = 5.9 GHz. (b) Transmission amplitude through a two cavity system with nonidentical qubits; $t_{c,L}$ = 21 $\mu$eV, $t_{c,R}$ = 16 $\mu$eV. For both plots t = 0.035 $\mu$ eV, $\epsilon$ = 0, $\gamma_c/2\pi$ = 90 MHz, $g_c/2\pi$ = 50 MHz, $\kappa_1/2\pi$ = $\kappa_2/2\pi$ = 0.9 MHz. The energy eigenvalues of $H_{ph,2}$ are plotted as dashed lines along the y-axis.}
        \label{fig:trans_2cavity2dot}
\end{figure}

The behavior of this two-cavity system can be seen in Fig.~\ref{fig:trans_2cavity2dot}. As expected, the presence of two cavities gives rise to two different photon modes resulting in a transmission spectrum split into two bands shifted away from the single-cavity resonant frequency. The new transmission bands can be found when the energy of the driving field matches one of the energy eigenvalues, $\lambda_i$, of the empty-cavity photon Hamiltonian of the two-cavity system.
\begin{equation}
    H_{ph,2} = a^{\dagger}_La_L\omega_{cL} + a^{\dagger}_Ra_R\omega_{cR} +t(a_L^\dagger a_R + a_R^\dagger a_L).
    \label{eq:Hph2}
\end{equation}
\noindent As expected, the magnitude of the shift from the single cavity frequency, scales linearly with the inter-cavity coupling $t$ for identical cavities.

The shift in the transmitted frequencies is accompanied by a shift in the Rabi splittings, which now occur at two distinct magnetic field values. Qubits interacting with these photon excitations must be tuned to energies $E_{1(2)} - E_0 = \lambda_i$. Analogously, the new resonant magnetic field condition is determined by
\begin{equation}
    B_z^{res,i} = \lambda_i\sqrt{1 - \frac{B_x^2}{\lambda_i - 4t_c^2}}.
    \label{eq:Bzrespm}
\end{equation}

Just like in the single cavity case, we can obtain the strength of these interactions by measuring $g_s$ from the width of the corresponding Rabi splittings.

Figure~\ref{fig:trans_2cavity2dot}(a) shows results for the resonant regime, where both cavity and qubit parameters are identical. In this case, although each qubit is strongly coupled to its own cavity photon, the system is in the effective Heisenberg regime that can be achieved for two different coupling regimes~\cite{Borjans_2019}. Notice, however, that the spin-photon coupling values in the two transmission bands are nearly identical but not exactly equal. The spin-photon coupling strength for the higher-frequency mode is $g_{s,coupled}\approx0.0221$ $\mu$eV while the value for the lower frequency mode differs slightly, $g_{s,coupled}\approx0.0217$ $\mu$eV. This is due in part to the dependence of the relevant dipole matrix element ($d_{01}$) on $B_z$ as well as qubit parameters $B_x$ and $t_c$ (see Appendix A for details). As the figure shows, the strong coupling regime is reached at different values of $B_z$ for each photon mode. Since $B_x$ and $t_c$ are fixed, the magnitude of the dipole moment matrix element differs slightly between the two modes at resonance.

Fig. ~\ref{fig:trans_2cavity2dot}(b) shows results for non-identical qubit parameters. Here, it is clear that each photon mode couples to both qubits, but at different values of $B_z$. For each mode, one qubit couples more strongly to the cavity than the other due to the differing values of $t_c$. For example, in the high frequency transmission band, the coupling strengths for the individual qubits are: $g_{s,R}\approx0.0155$ $\mu$eV and $g_{s,L}\approx0.0062$ $\mu$eV. This suggests the possibility of individual qubit tuning.

Finally, comparing the same frequency for both photon mode across panels (a) and (b) demonstrates the $\sqrt{N}$ scaling with the number of qubits simultaneously coupled to a given photon mode (as predicted by the Tavis-Cumming model mentioned in the previous section). Specifically, for the upper transmission band a measurement of the combined 2-qubit Rabi splitting produces: $g_{s,coupled}\approx0.0221$ $\mu$eV and $g_{s,R}\approx0.0155$ $\mu$eV for a single qubit, values that are consistent with the same scaling pattern to within $\sim$1\%.
Similarly, for the lower frequency transmission band, an identical measurement renders a combined Rabi splitting of $g_{s,coupled}\approx0.0217$ $\mu$eV, while in Fig.~\ref{fig:trans_2cavity2dot}(b), a measurement of the Rabi splitting for the isolated right qubit produces the value $g_{s,R}\approx0.0150$ $\mu$eV. These results exhibit the $\sqrt{N}$ scaling to within $\sim$2\%.

Similar behavior can be seen when both input and output ports are connected to the same cavity. Following the same procedure as before, we calculate the transmission amplitude through a type-B 2-cavity system and obtain:
\begin{equation}
    A_2^{(B)} = \frac{-i \xi_R \sqrt{\kappa_1\kappa_2}}{\xi_R (\xi_L+i\frac{\kappa}{2}) -t^2}
    \label{A_2alt}
\end{equation}

\begin{figure}
    \centering
    \includegraphics[width=0.48\textwidth]{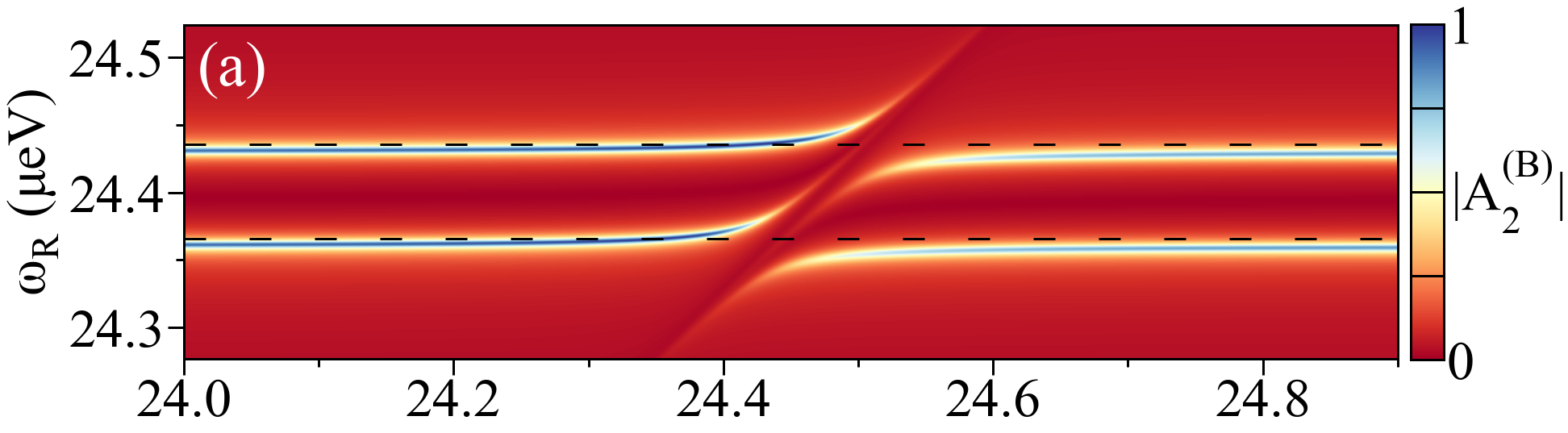}
    \hfill
    \includegraphics[width=0.48\textwidth]{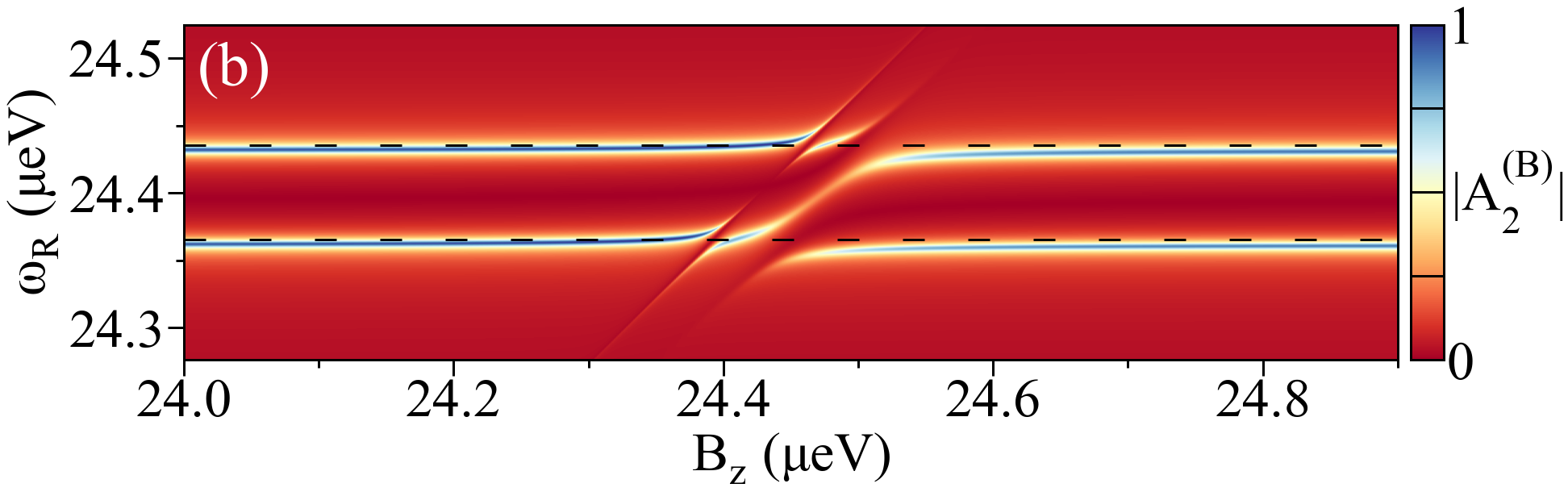}
    \hfill
    \includegraphics[width=0.48\textwidth]{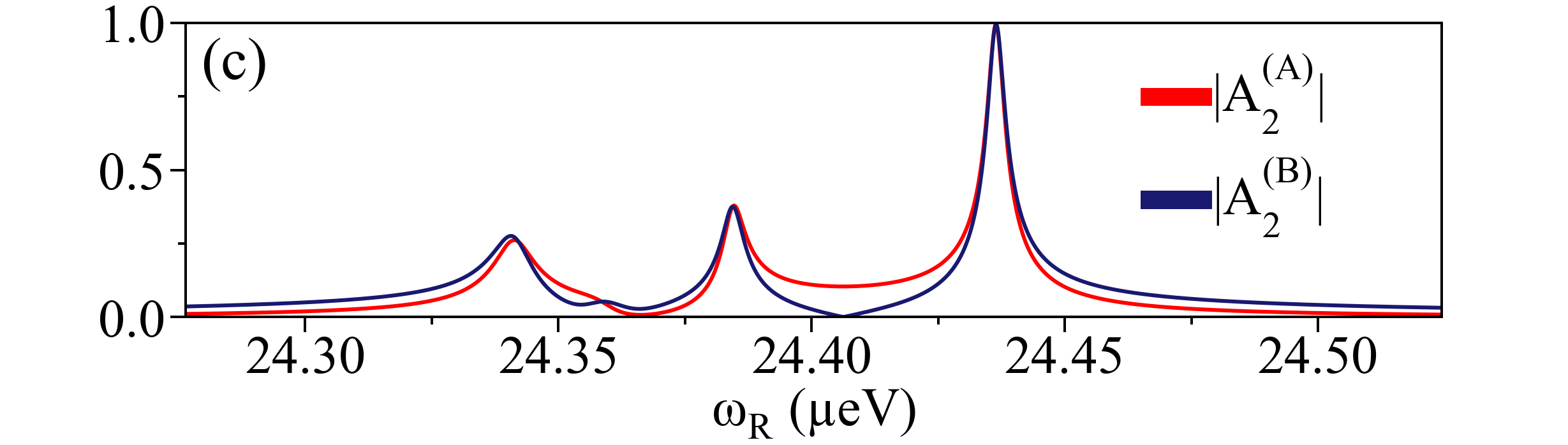}
        \caption{(a) Transmission amplitude through a type-B two-cavity, two-qubit system as a function of $B_z$ and $\omega_R$. In this plot the two qubits have identical parameters matching those in Fig.~\ref{fig:trans_2cavity2dot}(a). The energy eigenvalues of $H_{ph,2}$ are plotted as dashed lines along the y-axis. (b) Transmission amplitude through a type-B two-cavity system with nonidentical qubits; $t_{c,L}$ = 21 $\mu$eV, $t_{c,R}$ = 16 $\mu$eV. (c) Transmission amplitude vs driving frequency at the resonant magnetic field value $B_z^{res} \approx$ 24.44 $\mu$eV for the systems shown in Fig.~\ref{fig:trans_2cavity2dot}(a) and Fig.~\ref{fig:trans_2cavity2dot_alt}(a). This plot shows the difference in behavior between the type-A and type-B layouts for an otherwise identical system. For all plots t = 0.035 $\mu$eV, $\epsilon$ = 0, $\gamma_c/2\pi$ = 90 MHz, $g_c/2\pi$ = 50 MHz, $\kappa_1/2\pi$ = $\kappa_2/2\pi$ = 0.9 MHz, $\omega_{c,L}/2\pi$ = $\omega_{c,R}/2\pi$ = 5.9 GHz.}
        \label{fig:trans_2cavity2dot_alt}
\end{figure}

This expression confirms that the $R$ cavity in this setup does not contribute to photon dissipation as it is not connected to the external ports. The consequences of this equation can be seen in Fig.~\ref{fig:trans_2cavity2dot_alt}(a) where we plot the transmission amplitude through a type-B system with identical parameters to those of the type-A system shown in Fig.~\ref{fig:trans_2cavity2dot}(a). The transmission spectra of the two systems are largely consistent with one another, demonstrating the same shifting of the transmission bands, and the same $\sqrt{N}$ multi-qubit enhancement of the Rabi splittings. 
Panel~\ref{fig:trans_2cavity2dot_alt}(b) shows results for non-identical qubit parameters. As shown previously for the type-A layout, when tuned to different values of $t_c$, the individual qubits are in resonance with the photon modes at distinct values of $B_z$, and it is clear that the coupling strengths differ between qubits. For the high frequency transmission band, the coupling strengths of the qubits are: $g_{s,R}\approx0.0155$ $\mu$eV and $g_{s,L}\approx0.0060$ $\mu$eV, which demonstrate consistency with the analogous values for the type-A system to within $\sim$3\%.

In Fig.~\ref{fig:trans_2cavity2dot_alt}(c) we show a comparison between the two setups by plotting cuts from Fig.~\ref{fig:trans_2cavity2dot}(a) and Fig.~\ref{fig:trans_2cavity2dot_alt}(a) at a fixed value of the magnetic field $B_z$. In this plot we compare the strength of the Rabi splitting in the lower transmission band for both configurations. For the type-A system we obtain $g_{s,coupled}\approx0.0217$ $\mu$eV while for the type-B system we measure $g_{s,coupled}\approx0.0219$ $\mu$eV, showing a discrepancy within $\sim$1\%.

\subsubsection{Non-identical cavities}
\label{subsub1}
Since the fabrication of resonator cavities is never perfect, we consider the possibility that the two cavities have different resonant frequencies ($w_{c,L} \neq w_{c,R}$). Similarly to a change in the cavity coupling parameter $t$, a change in the cavity frequencies will lead to new photon eigenmodes and accompanying transmission bands that can be found by diagonalizing Eq.~(\ref{eq:Hph2}). In particular, with nonidentical cavities, the system can now support the excitation of non-symmetric photon modes, which, with a large enough difference in cavity frequencies $\Delta \omega_c = \omega_{c,R} - \omega_{c, L}$, can cause a decrease in the overall transmission of photons through the system. Since the decrease in amplitude occurs independently of the behavior of the qubits in the system, this effect can be studied by calculating the transmission through an empty multi-cavity system, i.e. containing zero qubits. Equations for the transmission amplitude through such zero-qubit 2-cavity systems for both schematic layouts are:
\begin{equation}
    A_2^{(A,0)} = \frac{-i t \sqrt{\kappa_1\kappa_2}}{(\Delta_L + i\frac{\kappa_1}{2}) (\Delta_R + i\frac{\kappa_2}{2}) - t^2}
    \label{eq:A20}
\end{equation}
\begin{equation}
    A_2^{(B,0)} = \frac{- i \Delta_R \sqrt{\kappa_1\kappa_2}}{\Delta_R(\Delta_L + i \frac{\kappa_1+\kappa_2}{2}) - t^2}
    \label{eq:A20alt}
\end{equation}

In these expressions, the value of the hopping parameter $t$ is determined by the resonant frequencies of the two individual cavities ($t \propto \omega_{c,L}\omega_{c,R}$)~\cite{PhysRevA.86.023837}. Because the bare frequencies are shifted when the cavities are connected, the value of $t$ changes, however, this is a second order effect and we take $t$ to be a constant parameter in the results presented below~\cite{PhysRevA.86.023837}.

\begin{figure}
    \centering
    \includegraphics[width=0.48\textwidth]{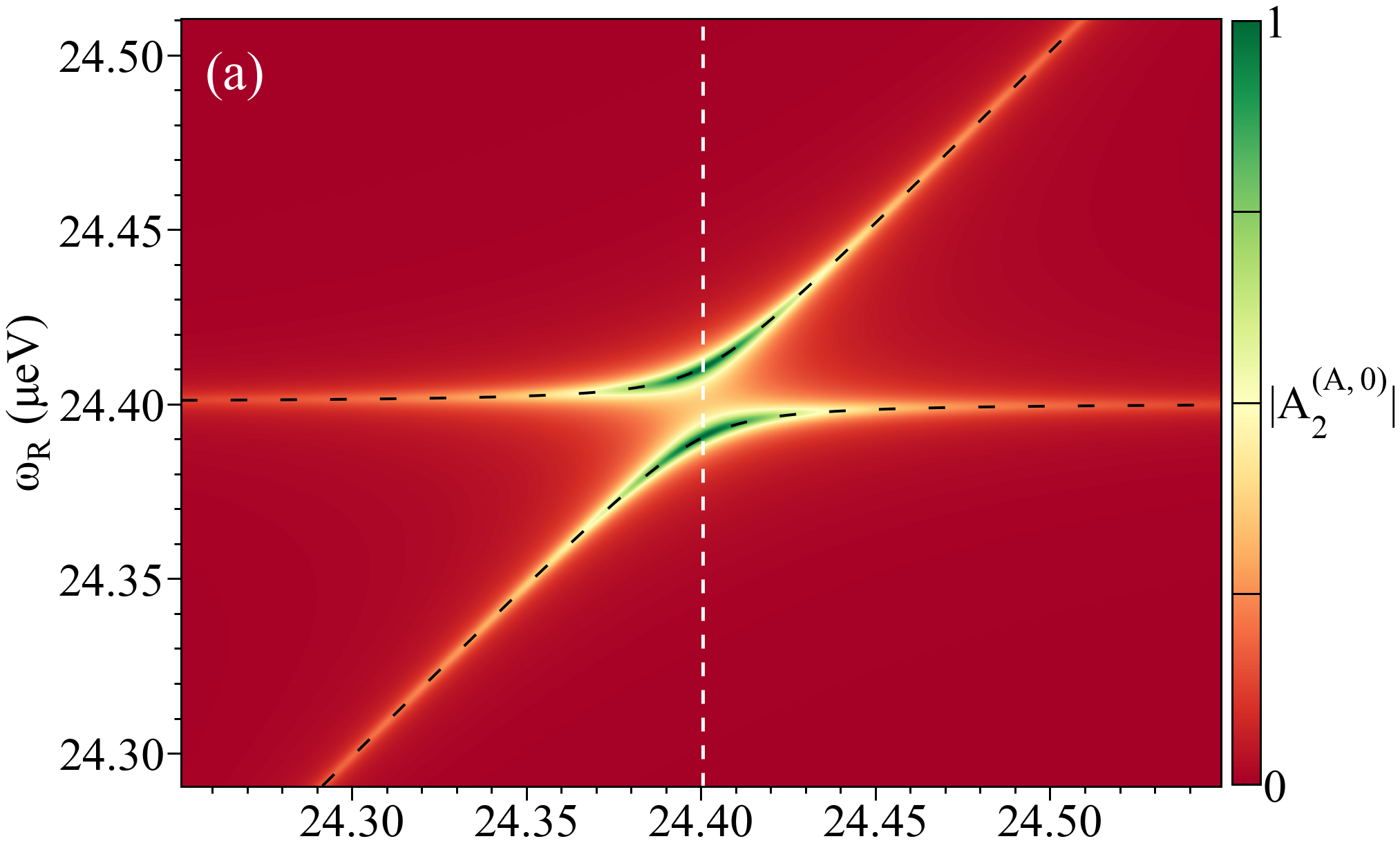}
    \hfill
    \includegraphics[width=0.48\textwidth]{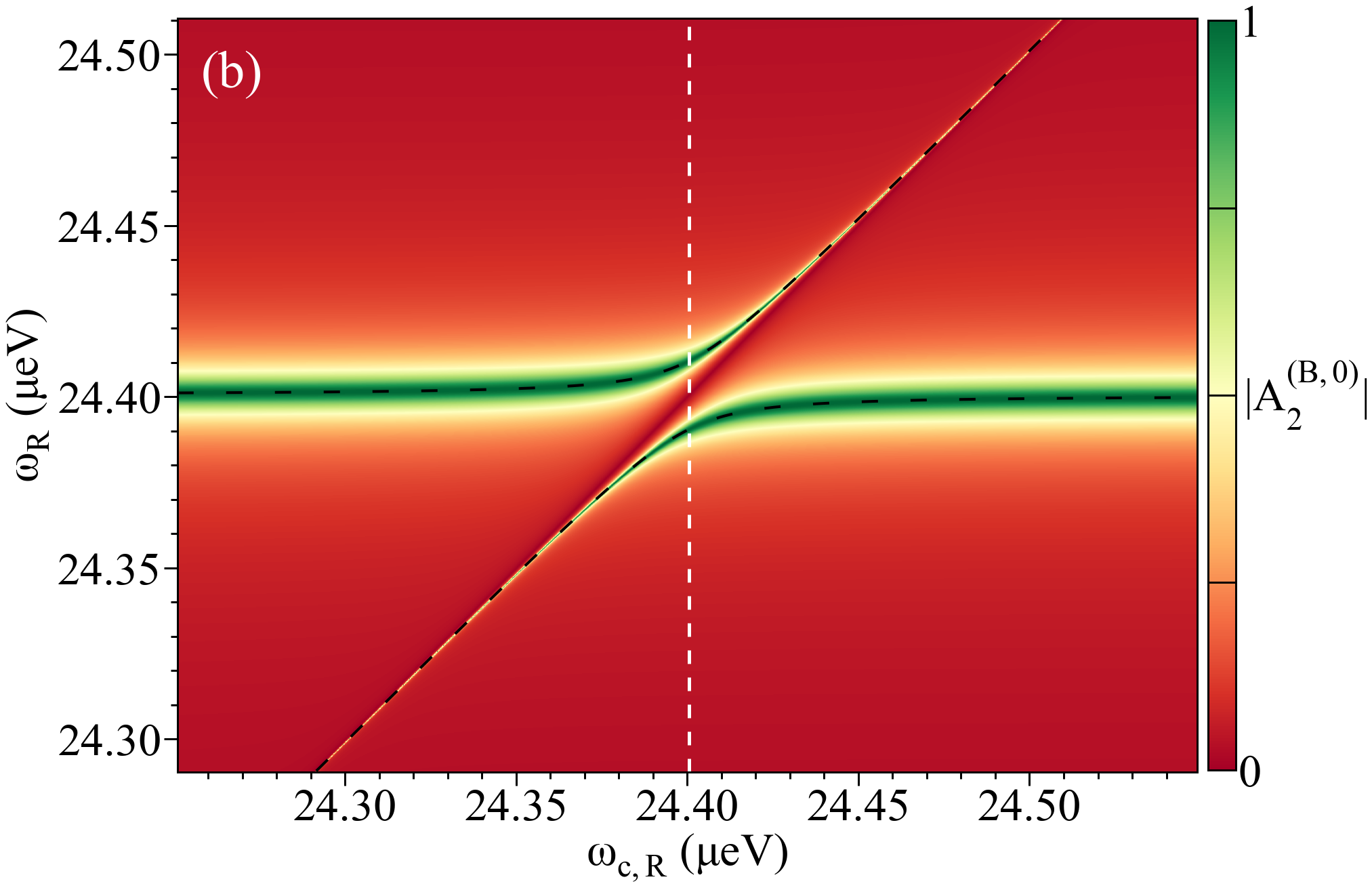}
        \caption{Transmission amplitude through (a) a type-A two-cavity system with 0 qubits and (b) a type-B two-cavity system with 0 qubits for varying mismatched values of the resonant cavity frequencies $\omega_{c,L}$ and $\omega_{c,R}$. Transmission is plotted against the driving frequency $\omega_R$ on the y-axis. In both plots $w_{c,L}$ is held constant at $\omega_{c,L}/2\pi$ = 5.9 GHz which is marked on the x-axis as a white vertical line. The resonant frequency of the right cavity is then allowed to vary along the x-axis. The energy eigenvalues of $H_{ph,2}$ are plotted as black dashed lines. In both plots $\kappa_1/2\pi$ = $\kappa_2/2\pi$ = 0.9 MHz, t = 0.01 $\mu$eV.}
        \label{fig:trans_2cavityempty}
\end{figure}

The amplitudes resulting from these equations are plotted for various mismatched values of resonant cavity frequencies in Fig.~\ref{fig:trans_2cavityempty}. Note that the transmission bands in both layouts are similar because  they track the same energy eigenvalues of $H_{ph,2}$.

As the gap $\Delta \omega_c$ between the cavity frequencies increases, the photon eigenmodes of the system move away from the symmetric modes with eigenvectors $[\frac{1}{\sqrt{2}}, \frac{1}{\sqrt{2}}]$ and $[\frac{1}{\sqrt{2}}, \frac{-1}{\sqrt{2}}]$,  towards the isolated modes $[1,0]$, $[0,1]$ (working in a basis where $[1,0]$ represents a photon in the left cavity and $[0,1]$ represents a photon in the right cavity). High transmission in the type-A system requires a significant photon population in both the input and output cavities; thus, as the system moves towards the isolated eigenmodes where the photon excitations are concentrated in just one cavity, the overall transmission dies out. This is evident in Fig.~\ref{fig:trans_2cavityempty}(a), where the transmission amplitude falls off in both transmission bands as $\omega_{c,R}$ moves away from $\omega_{c,L}$. This behavior is especially pronounced when the cavity coupling is small. We find that as long as $\Delta\omega_{c} < 2t$ the transmission remains above $\sim$90\% of an equivalent symmetric type-A system.

Fig.~\ref{fig:trans_2cavityempty}(b) shows how the type-B system behaves differently with non-identical cavities. In the type-B system, transmission decreases along only one direction in each transmission band. As an increased gap in cavity frequencies drives the system toward the isolated photon eigenmodes $[1,0]$ and $[0,1]$, transmission from input port to output port remains high for the mode in which the photon population is concentrated
mostly in the left cavity, since this cavity is connected to both ports. Although the transmission amplitude remains strong in this case, an increased gap renders a system of two disconnected cavities. This behavior is discussed below in more detail for the three-cavity system.

\subsection{The three cavity system}
\label{sec3B}

\begin{figure}[h]
     \centering
     \includegraphics[width=0.48\textwidth]{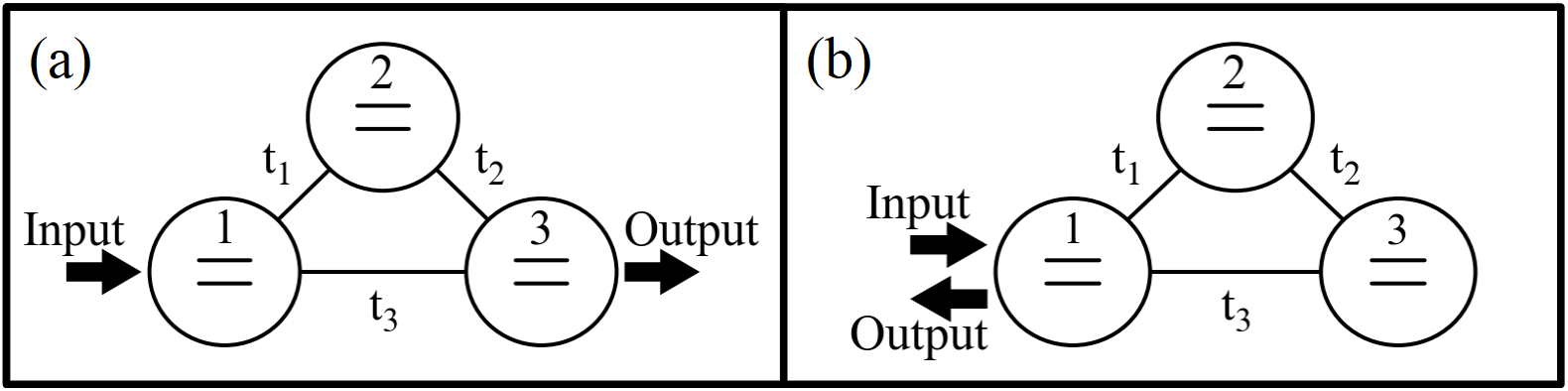}
     \hfill
        \caption{Schematic diagram showing two possible layouts for the three-cavity system. Three microwave resonator cavities (labeled 1, 2, 3), each containing a single DQD qubit, are connected via coupling capacitors with coupling strengths $t_1$, $t_2$, $t_3$. An input signal is sent in at cavity 1 and the output signal is measured from either (a) cavity 3 or (b) cavity 1.}
        \label{fig:3cavity_schematic}
\end{figure}

To explore two-dimensional connectivities, a natural extension of the model considers a three-cavity layout where all cavities are connected with coupling capacitors of strength $t_i$ as shown in Fig.\ref{fig:3cavity_schematic}. The photon exchange Hamiltonian for the three-cavity system describes the transfer of photons between any two cavities as:
\begin{dmath}
    H_{e,3} = t_1 a_1^{\dagger}a_2 + t_2a_2^{\dagger}a_3 + t_3a_1^{\dagger}a_3 + h.c.
    \label{eq:He3}
\end{dmath}
\noindent Thus the total three-cavity Hamiltonian is
\begin{equation}
    H_{sys,3} = H_1 + H_2 + H_3 + H_{e,3}
    \label{eq:Hsys3}
\end{equation}
\noindent where $H_{1,2,3}$ are the single cavity Hamiltonians for cavities 1,2 and 3 respectively.

As in the two-cavity case, there are two possible setups for the system. The type-A layout, where input and output ports are connected to separate cavities, and the type-B layout where a single cavity contains both ports. Configurations with different connectivities can be obtained by tuning the cavity coupling strengths $t_i$.

As in the previous section, we calculate the transmission amplitude. For a type-A three-cavity system the expression is:
\begin{equation}
    A_3^{(A)} = \frac{- i\sqrt{\kappa_1\kappa_2}(t_1 t_2 + t_3\xi_2)}{\xi_2\Bar{\xi}_1\Bar{\xi}_3 - t_2^2\Bar{\xi}_1 - t_1^2\Bar{\xi}_3 - t_3^2\xi_2 - 2t_1t_2t_3} 
    \label{eq:A3A}
\end{equation}
\noindent where $\Bar{\xi}_1 = \xi_1 +i\frac{\kappa_1}{2}$ and $\Bar{\xi}_3 = \xi_3 + i \frac{\kappa_2}{2}$.

As expected, the transmission spectrum splits into three transmission bands originating from the three photon modes arising from the coupled cavities. The normal mode frequencies can be found by diagonalizing the empty-cavities Hamiltonian given by:
\begin{dmath}
H_{ph,3} = \sum_{i=1}^3\omega_{c,i}a_i^{\dagger}a_i +  H_{e,3}
\end{dmath}

\begin{figure}
    \centering
    \includegraphics[width=0.48\textwidth]{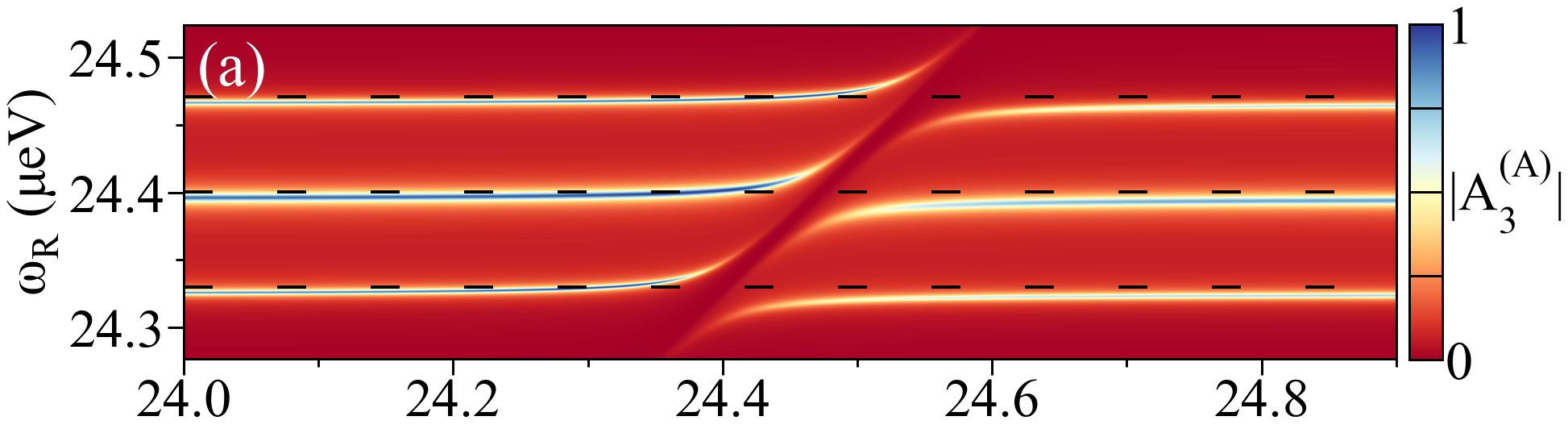}
    \hfill
    \includegraphics[width=0.48\textwidth]{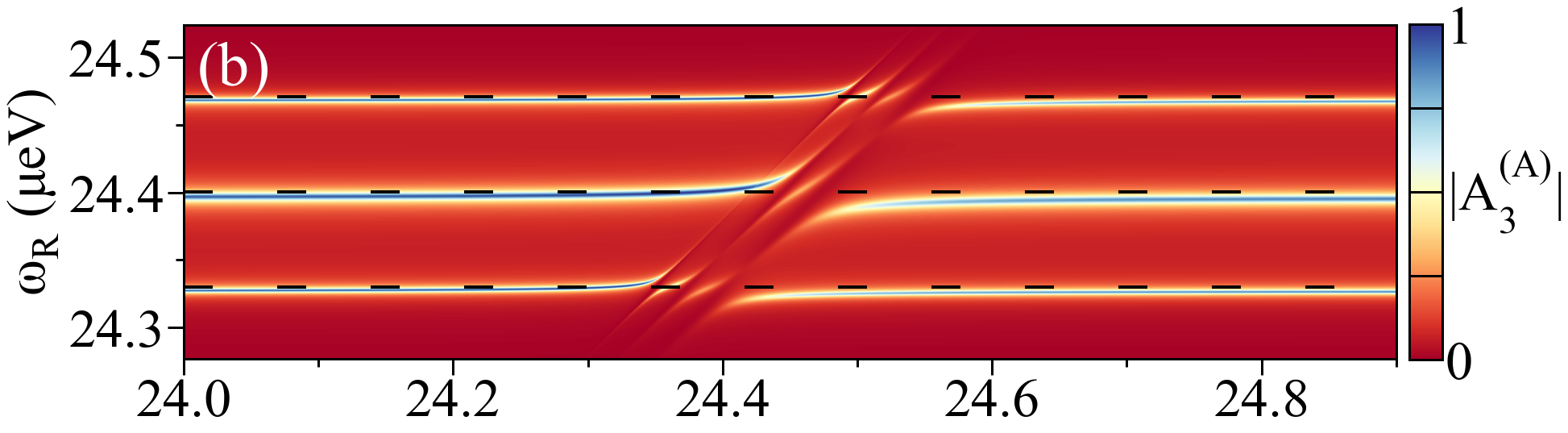}
    \hfill
    \includegraphics[width=0.48\textwidth]{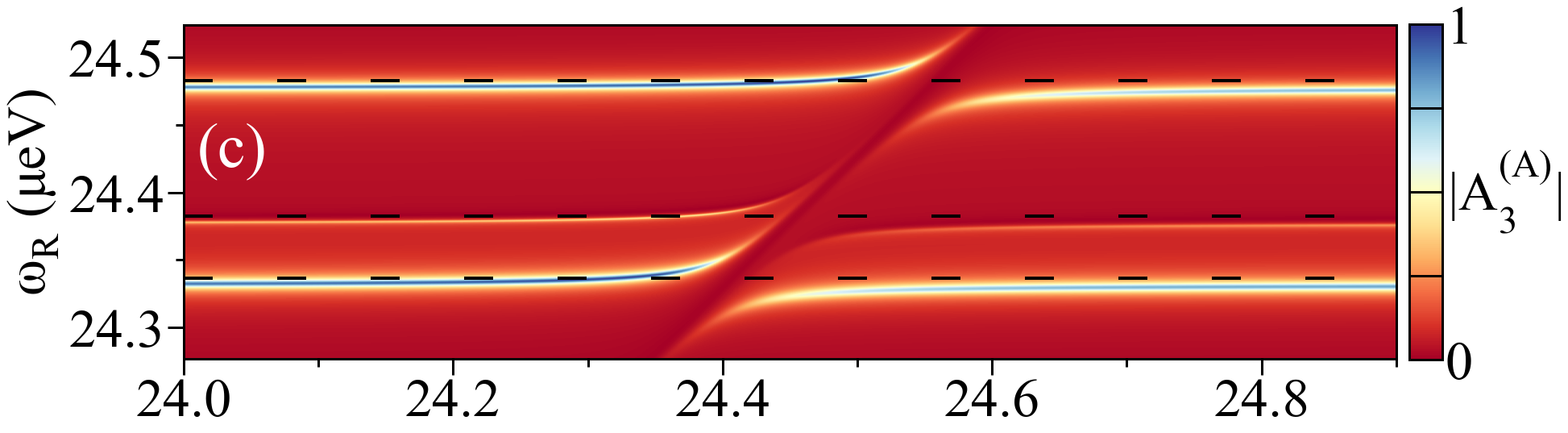}
    \hfill
    \includegraphics[width=0.48\textwidth]{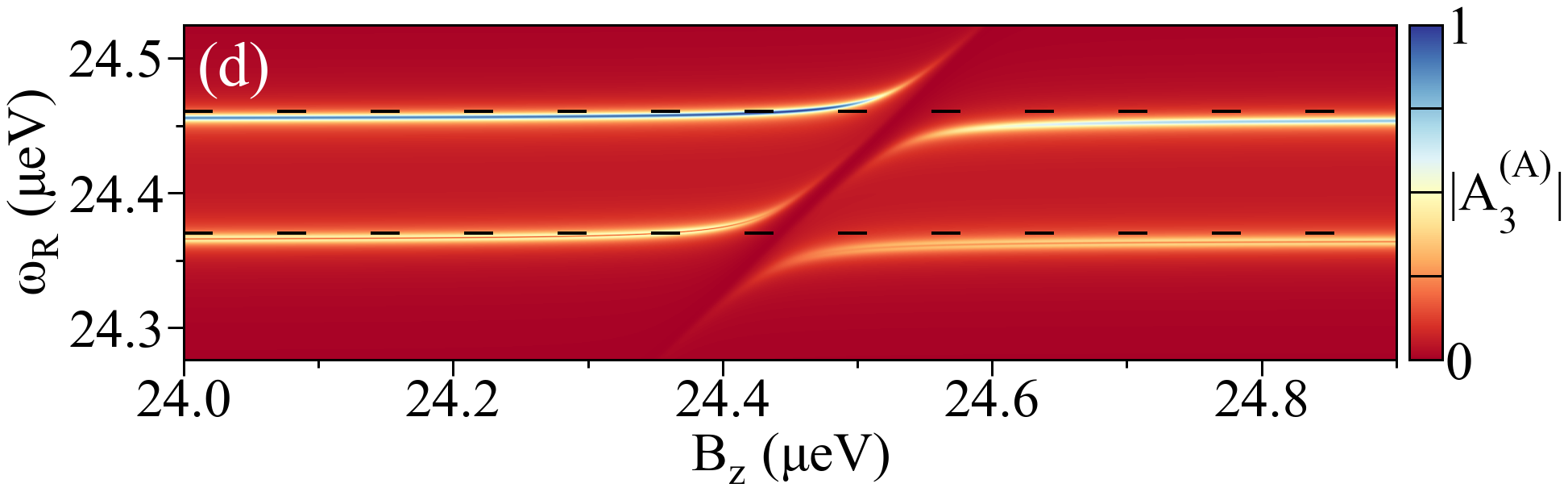}
        \caption{(a) Transmission amplitude through a type-A, three-cavity, three-qubit system as a function of $B_z$ and $\omega_R$. In this plot all qubits are identical with $B_{x,n}$=1.6 $\mu$eV, $t_{c,n}$ = 16 $\mu$eV, and all cavities are identical with $\omega_{c,n}/2\pi$ = 5.9 GHz. The three cavities are arranged in a linear chain with $t_1$ = $t_2$ = 0.05 $\mu$ eV and $t_3$ = 0 $\mu$ eV. (b) Transmission amplitude through the same system as in plot (a) but with nonidentical qubits where now $t_{c,1}$ = 16 $\mu$eV, $t_{c,2}$ = 23 $\mu$ eV, $t_{c,3}$ = 18 $\mu$eV. (c) Transmission amplitude through a three cavity system with identical qubits ($B_{x,n}$ = 1.6 $\mu$eV, $t_{c,n}$ = 16$\mu$eV), but with varying coupling strengths between cavities. Here, $t_1$ = 0.02 $\mu$eV, $t_2$ = 0.04 $\mu$eV, $t_3$ = 0.06 $\mu$eV. (d) Transmission amplitude through a three cavity system with identical qubits ($B_{x,n}$ = 1.6 $\mu$ eV, $t_{c,n}$ = 16$\mu$eV), and with identical coupling strengths for all connections, $t_1$ = $t_2$ = $t_3$ = 0.03$\mu$eV. For all plots $\epsilon$ = 0 $\mu$eV, $\gamma_c/2\pi$ = 90 MHz, $g_c/2\pi$ = 50 MHz, $\kappa_1/2\pi$ = $\kappa_2/2\pi$ = 0.9 MHz. The energy eigenvalues of $H_{ph,3}$ are plotted as dashed lines along the y-axis.}
        \label{fig:trans_3cavity3dot}
\end{figure}

In Fig.~\ref{fig:trans_3cavity3dot}, we show the transmission spectrum for a linear chain of 3 one-qubit cavities. The addition of a new cavity allows to explore a richer parameter space as shown by the different panels. For example, panel~(a) shows the resonant regime where cavities and qubits have identical parameters. Panel~(b) shows the regime with all identical cavities but different qubit parameters. In this case, the Rabi splittings for each qubit are seen separately. Notice that the top and bottom transmission bands present three distinct Rabi splittings, an indication that all three qubits are interacting with the coupled cavity system. In this case, the energy levels of these transmission bands match photon modes that exhibit a significant photon population in each cavity. The empty-cavity eigenmodes of $H_{ph,3}$ that correspond to these transmission bands are $[\frac{1}{2},\frac{1}{\sqrt{2}},\frac{1}{2}]$ and $[\frac{1}{2},\frac{-1}{\sqrt{2}},\frac{1}{2}]$ (here, we are working on the basis where $[1,0,0]$ represents a photon in cavity 1, $[0,1,0]$ represents a photon in cavity 2, and [0,0,1] represents a photon in cavity 3). The central transmission band, however, corresponds to the photon mode $[\frac{-1}{\sqrt{2}},0,\frac{1}{\sqrt{2}}]$ with no photon population in the central cavity. This explains the existence of only two Rabbi splittings in the central transmission band as the qubit in cavity 2 does not interact with the rest of the system.

Panel~(c) presents transmission data for asymmetric cavity couplings, showing a finite, albeit much reduced transmission at the central frequency.

Finally, in panel~(d) demonstrates what happens when the system is driven at a frequency corresponding to degenerate eigenmodes. Here, all three cavities are connected with identical couplings ($t_1=t_2=t_3$), and the photon normal modes [-0.82,0.41,0.41] and [0.29,-0.81,0.52] are degenerate. This produces a merging of   two distinct transmission bands which converge into one with reduced amplitude, signaling a destructive interference effect.

In general, the strength of the Rabi splitting for any individual qubit will scale proportionally to the square root of the number of photons in it's cavity (consistent with the Jaynes-Cummings model ~\cite{JC1963}). For example, in the high-frequency transmission band in Fig.~\ref{fig:trans_3cavity3dot}(c), which corresponds to the normal mode [0.59, 0.46, 0.66], we would expect to find $g_{s,1}/0.59= g_{s,2}/0.46= g_{s,3}/0.66= g_{s,coupled}$. When measured, these values are found to be in agreement with each other to within $\sim$1\%.

Next, we analyze  a three-cavity system where the input and output ports are connected to the same cavity. The transmission amplitude is:

\begin{equation}
    A_3^{(B)} = \frac{-i (\xi_2\xi_3-t_2^2)\sqrt{\kappa_1\kappa_2}}{\Bar{\xi}_{1,\kappa}\xi_2\xi_3 - t_2^2\Bar{\xi}_{1,\kappa} -t_1^2\xi_3-t_3^2\xi_2-2t_1t_2t_3}
    \label{eq:A3B}
\end{equation}

\noindent where $\Bar{\xi}_{1,\kappa} =\xi_1 +i\frac{\kappa}{2}$.

\begin{figure}
    \centering
    \includegraphics[width=0.48\textwidth]{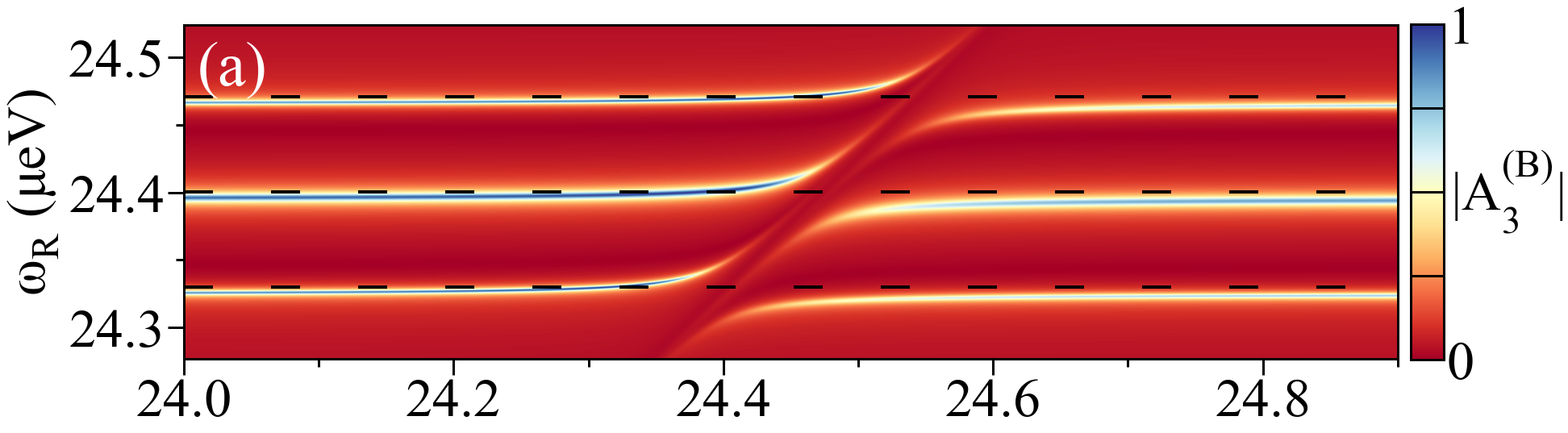}
    \hfill
    \includegraphics[width=0.48\textwidth]{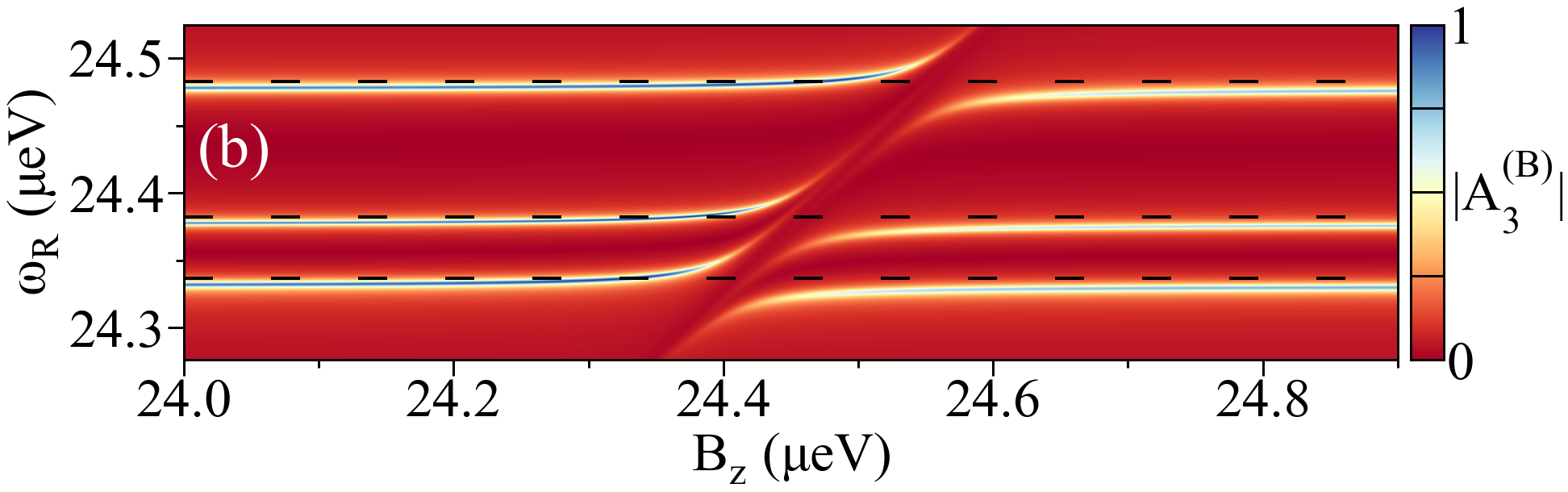}
    \hfill
    \includegraphics[width=0.48\textwidth]{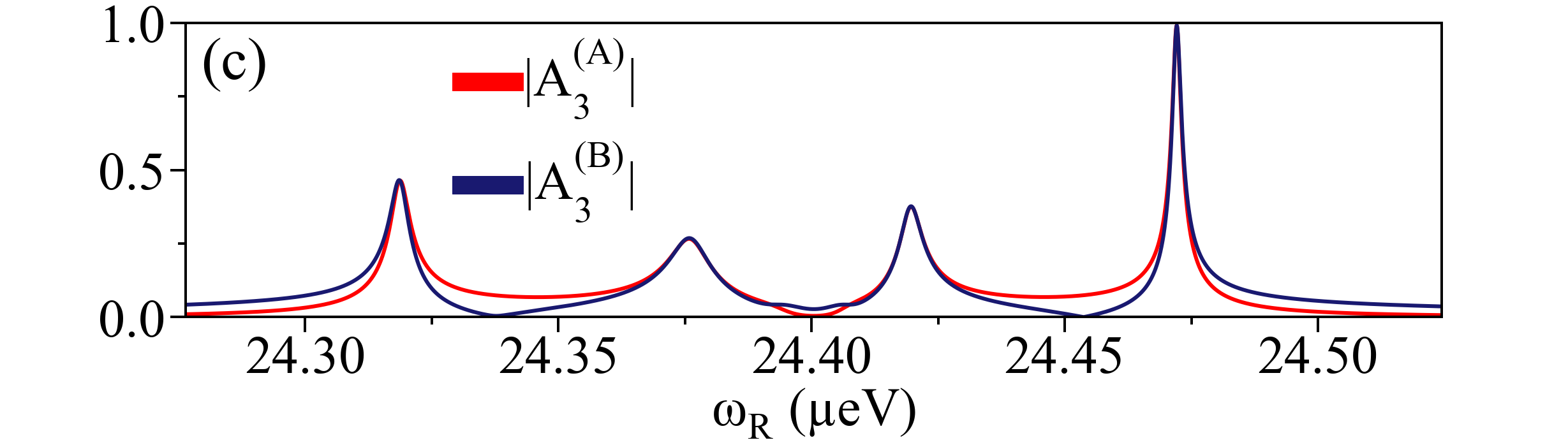}
        \caption{(a) Transmission amplitude through a type-B three-cavity, three-qubit system as a function of $B_z$ and $\omega_R$. In this plot all qubits and cavities are identical, with parameters matching those in Fig.~\ref{fig:trans_3cavity3dot}(a). The energy eigenvalues of $H_{ph,3}$ are plotted as dashed lines along the y-axis. (b) Transmission amplitude through a three cavity system with identical qubits, but with varying coupling strength between cavities. Here, all qubit and cavity parameters match those in Fig.~\ref{fig:trans_3cavity3dot}(c). (c) Transmission amplitude vs driving frequency at the resonant magnetic field value $B_z^{res} \approx$ 24.47 $\mu$eV for the linear chain systems shown in Fig.~\ref{fig:trans_3cavity3dot}(a) and Fig.~\ref{fig:trans_3cavity3dot_alt}(a). This plot shows the difference in behavior between the type-A and type-B layouts for an otherwise identical system. For all plots $\epsilon$ = 0 $\mu$eV, $\gamma_c/2\pi$ = 90 MHz, $g_c/2\pi$ = 50 MHz, $\kappa_1/2\pi$ = $\kappa_2/2\pi$ = 0.9 MHz.}
        \label{fig:trans_3cavity3dot_alt}
\end{figure}

In Fig.~\ref{fig:trans_3cavity3dot_alt} we show results for the same parameters used in Figs.~\ref{fig:trans_3cavity3dot}(a) and ~\ref{fig:trans_3cavity3dot}(c), with the only difference being the location of the output port. Fig.~\ref{fig:trans_3cavity3dot_alt}(c) shows cuts of the transmission spectra for type-A and type-B linear chain setups, emphasizing the differences between the systems away from the corresponding Rabi regimes.

One notable difference between the two setups appears when comparing the strengths of the central transmission bands for a given photon mode. This is clear when comparing Figures \ref{fig:trans_3cavity3dot}(c) and \ref{fig:trans_3cavity3dot_alt}(b). In this non-symmetric cavity coupling regime, the central transmission band corresponds to the photon mode [0.52, -0.84, 0.12]. This mode has a relatively low photon population in cavity 3, leading to low transmission through the type-A setup since the output of the system is measured off of this cavity. However, in the type-B setup, the output is measured off of 
cavity 1, and so the low photon population in cavity 3 does not cause an attenuation at this frequency.

As was done for the two-cavity system, we study the effects of mismatched resonant cavity frequencies by calculating the transmission amplitude through a system with zero qubits. For both setups type A and B, the transmission equation for the empty three cavity system can be obtained directly from Eqs.~(\ref{eq:A3A}-\ref{eq:A3B}) by replacing $\xi_i$ by $\Delta_i$. The  expression for the type-A setup is:
\begin{equation}
    A_3^{(A,0)} = \frac{- i\sqrt{\kappa_1\kappa_2}(t_1 t_2 + t_3\Delta_2)}{\Bar{\Delta}_{1,\kappa}\Bar\Delta_2{\Delta}_3 - t_2^2\Bar{\Delta}_{1,\kappa} - t_1^2\Bar{\Delta}_3 - t_3^2\Delta_2 - 2t_1t_2t_3} 
    \label{eq:A3A0}
\end{equation}
\noindent Here $\Bar{\Delta}_{1,\kappa} = \Delta_1+i\frac{\kappa_1}{2}$ and $\Bar{\Delta}_3 = \Delta_3 +i\frac{\kappa_2}{2}$. For the setup type-B we find:
\begin{equation}
    A_3^{(B, 0)} = \frac{-i (\Delta_2\Delta_3-t_2^2)\sqrt{\kappa_1\kappa_2}}{\Bar{\Delta}_{1,\kappa}\Delta_2\Delta_3 - t_2^2\Bar{\Delta}_{1,\kappa} -t_1^2\Delta_3-t_3^2\Delta_2-2t_1t_2t_3}
    \label{eq:A3B0}
\end{equation}
\noindent where $\Bar{\Delta}_{1,\kappa} = \Delta_1+i\frac{\kappa}{2}$.

\begin{figure}
    \centering
    \includegraphics[width=0.48\textwidth]{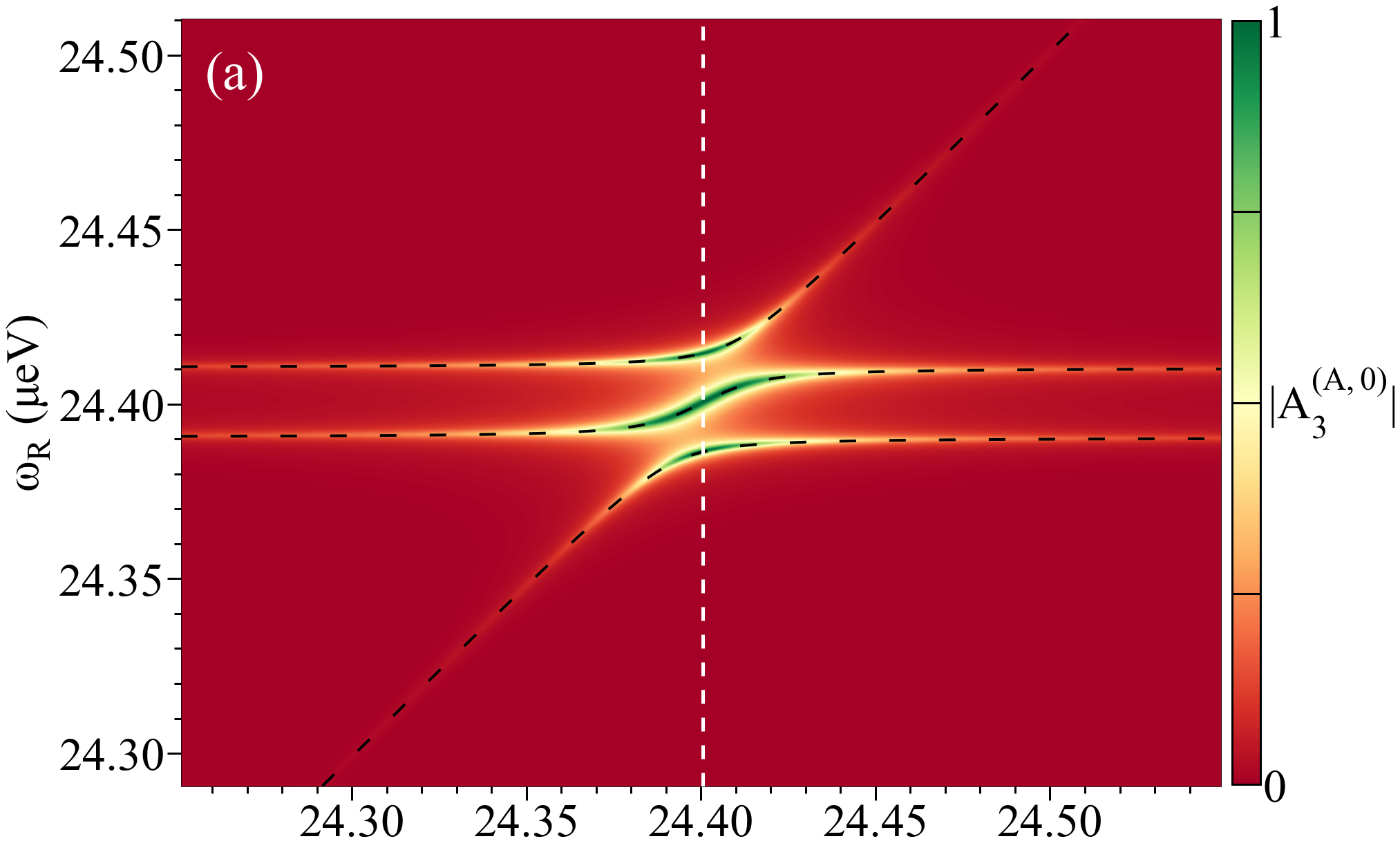}
    \hfill
    \includegraphics[width=0.48\textwidth]{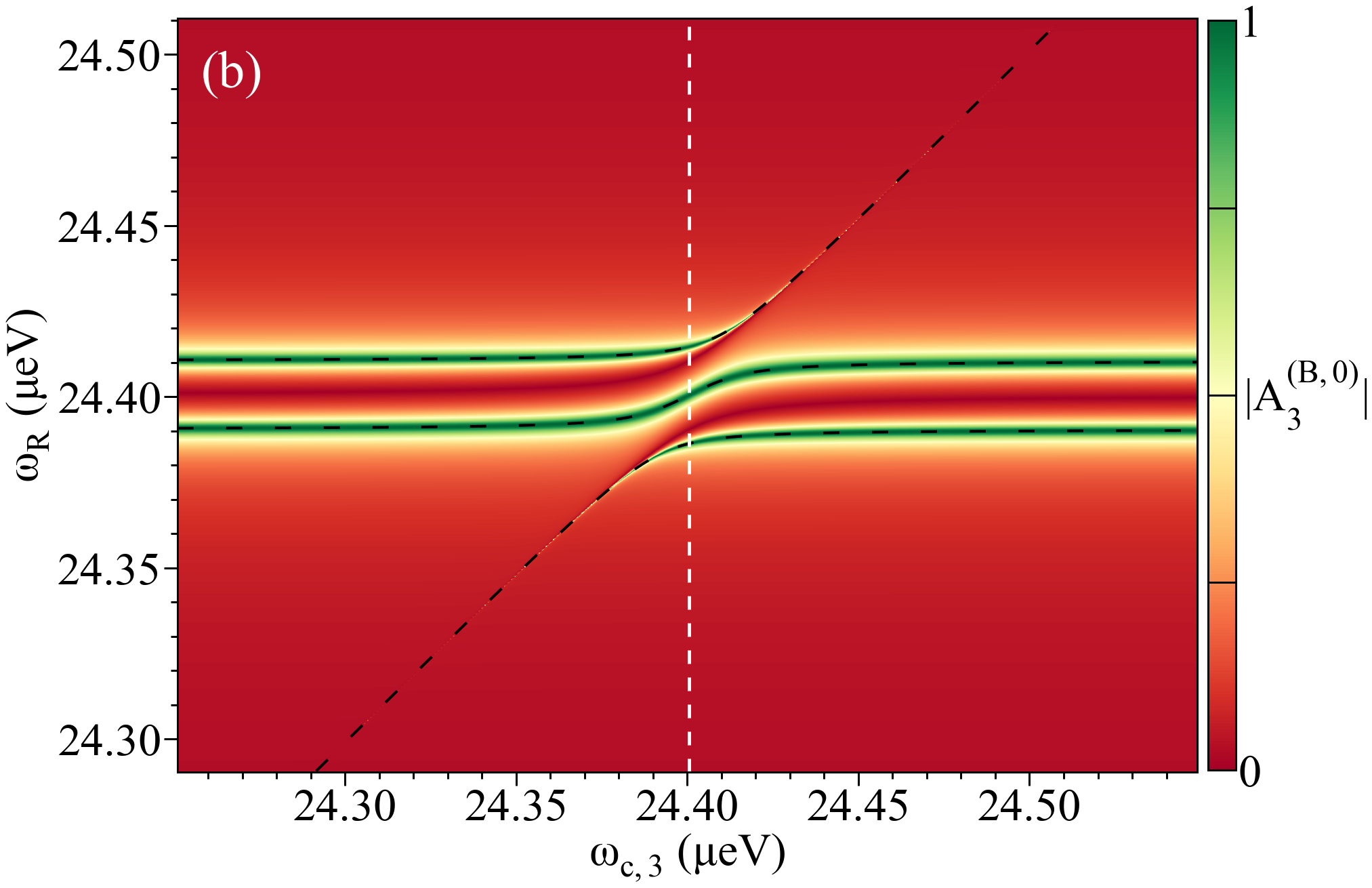}
        \caption{Transmission amplitude through three-empty cavities with (a) a type-A  and (b) a type-B configurations, for varying mismatched values of the isolated cavity frequencies $\omega_{c,1}$, $\omega_{c,2}$, and $\omega_{c,3}$. Transmission is plotted against the driving frequency $\omega_R$ on the y-axis. In both plots $\omega_{c,1}$ and $\omega_{c,2}$ are held constant at $\omega_{c,1}/2\pi$ = $\omega_{c,2}/2\pi$ = 5.9 GHz as marked on the x-axis with a white vertical line. The resonant frequency of cavity 3 is then allowed to vary. The energy eigenvalues of $H_{ph,3}$ are plotted as black dashed lines. In both plots $\kappa_1/2\pi$ = $\kappa_2/2\pi$ = 0.9 MHz, $t_1$ = 0.01 $\mu$eV, $t_2$ = 0.01 $\mu$eV, $t_3$ = 0 $\mu$eV.}
        \label{fig:trans_3cavityempty}
\end{figure}

Fig.~\ref{fig:trans_3cavityempty} shows the transmission amplitudes $A_3^{(A, 0)}$ and $A_3^{(B, 0)}$ calculated for a linear chain ($t_3=0$), with nonidentical cavities. As the cavity frequencies shift away from one another, there is a decrease in transmission through the type-A setup across all transmission bands, just as in the two cavity case. The type-B setup allows transmission only along the bands that correspond to modes with high photon populations in cavity 1, which is connected to the external ports. However, as before, the bands that retain high transmission even at highly mismatched cavity frequencies correspond to modes with low photon populations in cavity 3, meaning that the cavity is effectively cut off from the rest of the system.

\section{Hybrid Systems}
\label{sec4}

The model allows for extensions to hybrid systems that combine the single-cavity approach of placing multiple qubits inside a single cavity, with the multi-cavity layouts described above. In this section we discuss, as an example, a hybrid system comprised of two connected cavities, with $N_R$ and $N_L$ number of qubits in the right and left cavities, connected to output ports in a type-A layout. Here, qubits are labeled $q_{mn}$ where $m$ is the index of the cavity, and $n$ is the index of the qubit within the cavity.

For this hybrid system, Eq.~(\ref{eq:A2}) can be modified to include additional qubits to render a transmission given by:
\begin{equation}
    A_2^{(A, N_L,N_R)} = \frac{- i t \sqrt{\kappa_1\kappa_2}}{(\xi_L^{(N_L)} + i \frac{\kappa_1}{2})(\xi_R^{(N_R)}+i\frac{\kappa_2}{2}) - t^2}
    \label{eq:A2hybrid}
\end{equation}
\noindent where $\xi_i^{(N)}$ includes a sum over the $N$ qubits in cavity $i$:
\begin{equation}
    \xi_i^{(N)} =\Delta_i -\sum_{n=1}^Ng_{c,i,n}(d_{01,i,n}\chi_{01,i,n} + d_{02,i,n}\chi_{02,i,n}).
    \label{eq:xiN}
\end{equation}

The amplitudes for the type-B setup can be obtained by replacing Eq.~(\ref{eq:xiN}) in Eq.~(\ref{A_2alt}). Furthermore, the same substitution can be used to obtain transmission amplitudes through the three-cavity system in the two setups discussed.

\begin{figure}
    \centering
    \includegraphics[width=0.48\textwidth]{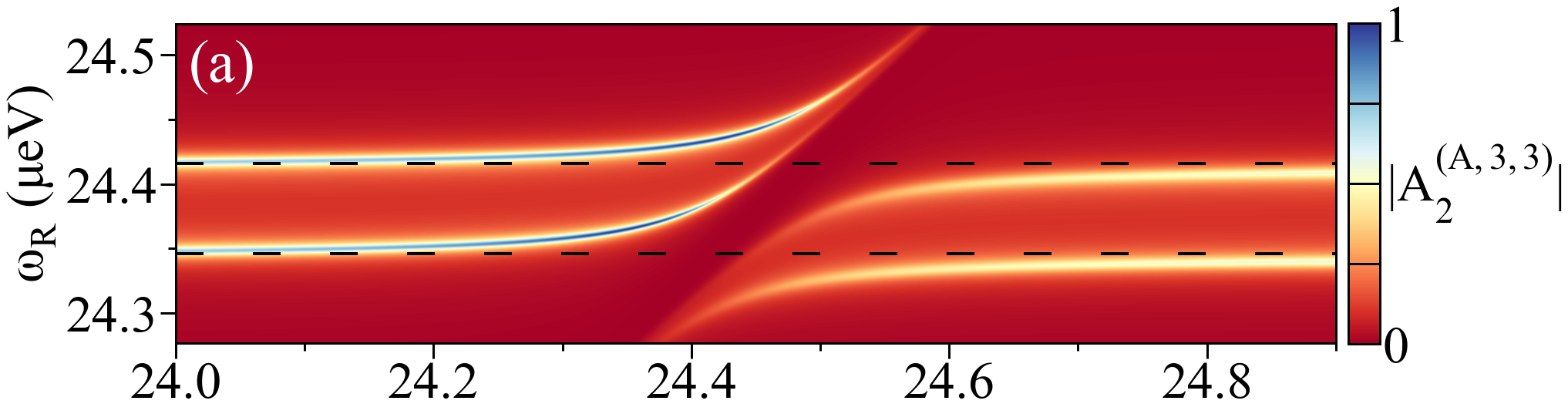}
    \hfill
    \includegraphics[width=0.48\textwidth]{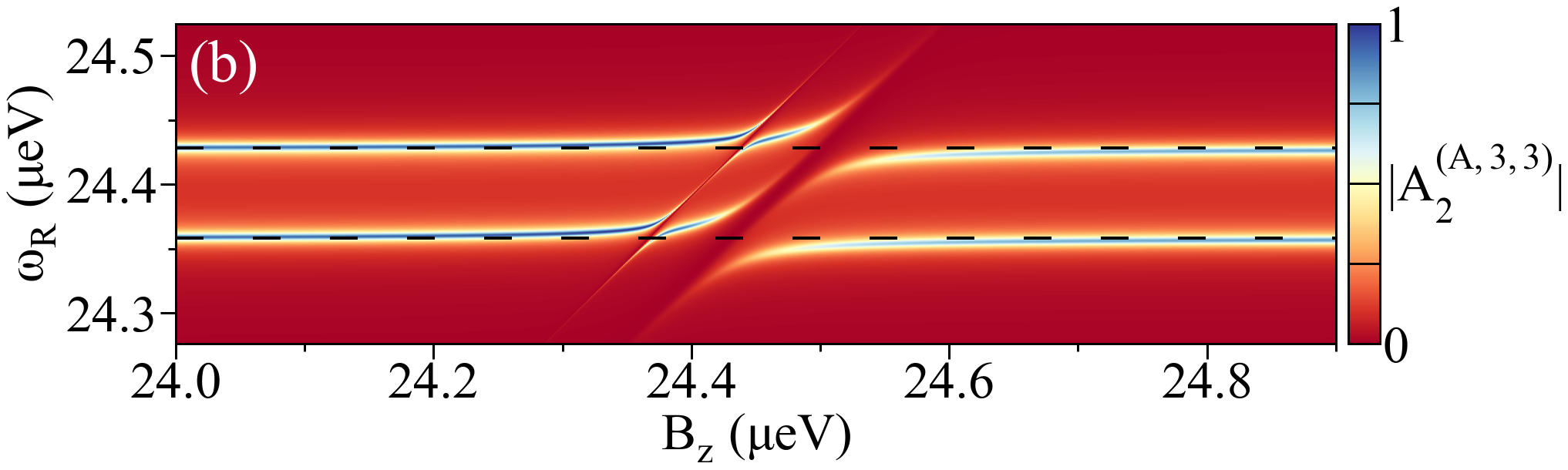} 
        \caption{(a) Transmission amplitude through a type-A 2-cavity system with 3 qubits in each cavity where all qubits are tuned in resonance with each other. Here, all qubits have identical parameters: $\epsilon$ = 0 $\mu$eV, $B_x$ = 1.6 $\mu$eV, $t_{c}$ = 16 $\mu$eV, $g_c/2\pi$ = 50 MHz. (b) Transmission amplitude through the same system as in subplot (a) but with 4 qubits shifted out of resonance by changing their tunnel coupling parameter $t_c$ such that $t_{c,L,2} = t_{c,L,3} = t_{c,R,2} = t_{c,L,3}$ = 24 $\mu$eV. All other qubit parameters are left unchanged. For both subplots $\omega_{c,L}/2\pi$ = $\omega_{c,R}/2\pi$ = 5.9 GHz, $\kappa_1/2\pi$ = $\kappa_2/2\pi$ = 0.9 MHz, $\gamma_c/2\pi$ = 90 MHz, t = 0.035 $\mu$eV.}
        \label{fig:trans_hybrid}
\end{figure}

Results obtained with Eq.~(\ref{eq:A2hybrid}), are shown in Fig.~\ref{fig:trans_hybrid} for two coupled cavities containing 3 qubits each. In panel~\ref{fig:trans_hybrid}(a) all qubits are tuned to be in simultaneous resonance (with identical tunnel couplings $t_{c}$ = 16 $\mu$eV), and the transmission spectrum exhibits an enhanced Rabi splitting in both transmission bands. In the high frequency band, the strength of this splitting is $g_{s,6}\approx 0.0379$ $\mu$eV. Panel ~\ref{fig:trans_hybrid}(b) shows a regime where 2 selected qubits have been isolated, so that their interaction can be distinguished from the rest of the system. This capability is crucial for implementing multi-qubit logic gates that leave the states of the other qubits unaffected. The two distinct Rabi splittings shown in each transmission band correspond to: 1) qubits $q_{L1}$ and $q_{R1}$ with identical tunnel couplings (the splitting shown on the right side of the panel) , and 2) qubits $q_{L2}$, $q_{L3}$, $q_{R2}$, and $q_{R3}$, which have been tuned to interact with the photon field at a distinct $B_Z$ value (the splitting shown on the left side of the panel). In this last case, the values for $B_Z$ have been produced by shifting the tunnel couplings to $t_c$ = 32 $\mu$eV. In the top transmission band, the strength of the Rabi splitting of qubits $q_{L1}$ and $q_{R1}$ is $g_{s,2}\approx 0.0219$ $\mu$eV. When compared, the measured values $g_{s,2}$ and $g_{s,6}$ are consistent with the expected $\sqrt{N}$ scaling to within $\sim$0.1\%.

\section{Conclusions}
\label{sec5}

Motivated by the need for scalable multi-qubit connectivity in spin-qubit platforms, we have extended the photon-coupled double-quantum-dot spin-qubit model of Ref.~\cite{Benito_2017} to architectures comprising two and three coupled cavities hosting single or multiple qubits. Such networks of coupled multi-qubit cavities enable higher qubit densities, while pushing scalibility beyond what any single-cavity can support. Using input-output theory, we first computed the photon transmission spectrum through a single cavity as a function of qubit number and found that the transmission signal is dramatically suppressed as additional qubits load the cavity — a limitation that becomes increasingly severe for larger qubit registers and directly constrains the scalability of shared-cavity architectures. Extending the model to multiple coupled cavities resolves this bottleneck: the transmission spectrum splits into distinct bands associated with different multi-cavity photon excitations, and the resulting frequency shifts allow the system to be driven away from the resonant single-cavity frequency.This tunability offers a practical route to circumventing specific frequency-dependent noise sources. 
We further show that the configuration of input/output ports plays a critical role in multi-cavity setups: the analytical expressions derived here enable straightforward identification of the parameters governing individual versus collective spin control, and we find that connecting external ports to a single, common cavity yields more robust overall transmission, particularly for non-identical cavities. Finally, we confirm that spin-photon coupling in multi-qubit, multi-cavity systems scales as $\sqrt{N}$ with qubit number, consistent with previous results for a single two-qubit cavity~\cite{Borjans_2019}, extending this scaling behavior to the modular multi-cavity setting for the first time. Taken together, these results establish coupled multi-cavity architectures as a promising and experimentally accessible strategy for scaling cQED-based spin-qubit platforms beyond the limitations of single-cavity designs.

\begin{acknowledgments}
We thank J. Petta, G. Burkard, M. Eriksson and M. Benito for useful discussions. We acknowledge the support of Army Research Office grant W911NF-23-1-0104.
\end{acknowledgments}

\appendix

\section{Single DQD-Cavity System}
\label{AppA}

Below we review the derivation of some results from Ref.~\cite{Benito_2017} that are used in the main text.

The basis is defined by $\{\ket{10\uparrow},\ket{01\uparrow},\ket{10\downarrow}\ket{01\downarrow}\}$, where (1,0) and (0,1) represent the charge states in the left and right quantum dots, and $\uparrow,  \downarrow$ represent the spin state in the $z$ direction. The Hamiltonian for the DQD qubit from Eq.~\ref{eq:Hb} is:
\begin{equation}
H_b = \frac{1}{2}
\begin{pmatrix}
    \epsilon + B_z & 2t_c & B_x & 0\\
    2t_c & -\epsilon + B_z & 0 & -B_x\\
    B_x & 0 & \epsilon - B_z & 2t_c\\
    0 & -B_x & 2t_c & -\epsilon - B_z
\end{pmatrix}.
\end{equation}

Introducing the parameter $\theta = \arctan{\frac{\epsilon}{2t_c}}$ we transform to a new basis $\{\ket{+\uparrow},\ket{-\uparrow},\ket{+\downarrow},\ket{-\downarrow}\}$ with states defined as
\begin{eqnarray*}
    \ket{+,\uparrow} &=& \frac{1}{\sqrt{2}}\left[\left( \cos{\frac{\theta}{2}} + \sin{\frac{\theta}{2}} \right) \ket{10} + \left( \cos{\frac{\theta}{2}} - \sin{\frac{\theta}{2}}\right)\ket{01}\right]  \\
    \ket{-,\uparrow} &=& \frac{1}{\sqrt{2}}\left[\left(\cos{\frac{\theta}{2}} - \sin{\frac{\theta}{2}}\right)\ket{10} - \left(\cos{\frac{\theta}{2} + \sin{\frac{\theta}{2}}}\right)\ket{01}\right]  \\
\ket{+,\downarrow} &=& \frac{1}{\sqrt{2}}\left[\left( \cos{\frac{\theta}{2}} + \sin{\frac{\theta}{2}} \right) \ket{10} + \left( \cos{\frac{\theta}{2}} - \sin{\frac{\theta}{2}}\right)\ket{01}\right]  \\
    \ket{-,\downarrow} &=& \frac{1}{\sqrt{2}}\left[\left(\cos{\frac{\theta}{2}} - \sin{\frac{\theta}{2}}\right)\ket{10} -  \left(\cos{\frac{\theta}{2} + \sin{\frac{\theta}{2}}}\right)\ket{01}\right] 
\end{eqnarray*}
\noindent In this basis the qubit Hamiltonian becomes
\begin{equation*}
H_b = \frac{1}{2}
\begin{pmatrix}
    \Omega + B_z & 0 & B_x\sin{\theta} & B_x\cos{\theta}\\
    0 & -\Omega + B_z & B_x\cos{\theta} & -B_x\sin{\theta}\\
    B_x\sin{\theta} & B_x\cos{\theta} & \Omega - B_z & 0\\
    B_x\cos{\theta} & -B_x\sin{\theta} & 0 & -\Omega - B_z
\end{pmatrix}
\end{equation*}
\noindent where $\Omega = \sqrt{\epsilon^2 + 4t_c^2}$.

When the system is set up with zero detuning ($\epsilon = 0$) then $\theta=0$ and the Hamiltonian reduces to
\begin{equation}
H_b^{(\epsilon = 0)} = \frac{r}{2}
\begin{pmatrix}
    \frac{2|t_c| + B_z}{r} & 0 & 0 & \sin{\phi}\\
    0 & -\cos{\phi} & \sin{\phi} & 0\\
    0 & \sin{\phi} & \cos{\phi} & 0\\
    \sin{\phi} & 0 & 0 & \frac{-(2|t_c| + B_z)}{r}
\end{pmatrix}
\end{equation}
\noindent where $\phi = \arctan{\frac{B_x}{2|t_c| s - B_z}}$ and $r = \sqrt{(2t_c - B_z)^2 + B_x^2}$. 

The eigenstates of the DQD qubit system at zero detuning can be parameterized by $\tan\beta = \frac{B_x}{2t_c + B_z}$, resulting in the following expressions:
\begin{eqnarray*}
    \ket{0} &=& \sin{\frac{\beta}{2}}\ket{+\uparrow} - \cos{\frac{\beta}{2}}\ket{-\downarrow}  \\
    \ket{1} &=& \cos{\frac{\phi}{2}}\ket{-\uparrow} + \sin{\frac{\phi}{2}}\ket{+\downarrow} \\
    \ket{2} &=& \sin{\frac{\phi}{2}}\ket{-\uparrow} - \cos{\frac{\phi}{2}}\ket{+\downarrow}  \\
    \ket{3} &=& \cos{\frac{\beta}{2}}\ket{+\uparrow} + \sin{\frac{\beta}{2}}\ket{-\downarrow}
\end{eqnarray*}
\noindent with associated energy levels
\begin{eqnarray*}
    E_{3,0} &=& \pm\frac{1}{2}\sqrt{(2t_c - B_z)^2 + B_x^2}  \\
    E_{2,1} &=& \pm\frac{1}{2}\sqrt{(2t_c + B_z)^2 + B_x^2}.
\end{eqnarray*}
When $\beta$ is sufficiently small, the states $\ket{0}$ and $\ket{3}$ simplify to
\begin{equation}
    \ket{0} \approx \ket{-\downarrow} \, \,\,\,\,\,\,\,\,\,\, \ket{3} \approx \ket{+\uparrow}.
\end{equation}

With the system prepared in the ground state, a spin flip will correspond to transitions from $\ket{0}$ to either the $\ket{1}$ or $\ket{2}$ state depending on the value of $\phi$. Generally, when $2t_c>B_z$ the relevant transition is $0\leftrightarrow1$ and when $2t_c<B_z$, the transition is $0\leftrightarrow2$.

The dipole operator in this eigenbasis is written as:
\begin{equation}
d_{(\epsilon=0)} =
\begin{pmatrix}
    0 & d_{01} & d_{02} & 0\\
    d_{01} & 0 & 0 & d_{13}\\
    d_{02} & 0 & 0 & d_{23}\\
    0 & d_{13} & d_{23} & 0\\
\end{pmatrix}
\end{equation}
\noindent where
\begin{eqnarray}
    d_{01} &=& d_{23} = \sin{\Big(\frac{\beta + \phi}{2}}\Big) \nonumber \\
    d_{02} &=& -d_{13} = \cos{\Big(\frac{\beta + \phi}{2}}\Big)
\end{eqnarray}
From these dipole elements the effective spin-photon coupling strength can be estimated as $g_s = g_c|d_{01(2)}|$. As derived in Refs.~\cite{Benito_2017,Mielke2021}, the 
susceptibilities $\chi_{01(2)}$ are calculated from the stationary limit of the main text Eqs.~(\ref{eq:abardot})-(\ref{eq:sigmabardot}) as:
\begin{eqnarray}
    \chi_{01} &=& \frac{g_c\cos{\theta}\sin{\frac{\phi}{2}}}{\delta_1-i\gamma_{eff}^{(2)}} \nonumber \\
    \chi_{02} &=& -\frac{g_c\cos{\theta}\cos{\frac{\phi}{2}}}{\delta_2-i\gamma_{eff}^{(1)}}
\end{eqnarray}

where the effective qubit dephasing rate is 

\begin{equation}
    \gamma_{eff}^{(n)} = \frac{\gamma_c}{\delta_{n}}\left(\delta_2\sin^2{\frac{\phi}{2}}+\delta_1\cos^2{\frac{\phi}{2}}\right),
    \label{eq:gammaeff}
\end{equation}

\noindent and with detunings $\delta_n = E_n - E_0 - \omega_R$. In the regime where spin transitions are dominant , the spin decoherence rate can be estimated as $\gamma_s=\gamma_{eff}^{(2)}$ (in the $0\leftrightarrow1$ regime) or $\gamma_s=\gamma_{eff}^{(1)}$ (in the $0\leftrightarrow2$ regime). We refer the reader to Ref.~\cite{Benito_2017,Mielke2021} for more details.

\section{Equations of motion}
\label{appB}

In this section we present the final equations of motion obtained in the rotating wave approximation, used to calculate the transmission amplitudes for the two- and three-coupled cavities systems.  For all of the different multi-cavity configurations considered in the main text, the equations of motion for the qubit operators $\sigma_{i}$, remain unchanged from Eq.~\ref{eq:sigmabardot} since the qubits interact directly with the photons in their own cavity. Meaningful differences arise when calculating equations of motion for the photon operators which include new terms for cavity hopping as well as interactions with the external fields.

For the type-A setup of the two-cavity system, the relevant equations of motion are:
\begin{eqnarray*}
\Bar{\dot{a}}_{L}&=& (i\omega_{c,L} - i\omega_R + \frac{\kappa_1}{2})\bar{a}_L +
\\ && ig_c(d_{01,L}\bar{\sigma}_{01,L} + d_{02,L}\bar{\sigma}_{02,L}) + it\bar{a}_R - \sqrt{\kappa_1}\bar{b}_{in,1}\\
\Bar{\dot{a}}_{R} &=& (i\omega_{c,R} - i\omega_R + \frac{\kappa_2}{2})\bar{a}_R + \\&& ig_c(d_{01,R}\bar{\sigma}_{01,R} + d_{02,R}\bar{\sigma}_{02,R}) + it\bar{a}_L
\end{eqnarray*}

For the type-B setup, the corresponding equations are:
\begin{eqnarray*}
    \Bar{\dot{a}}_{L} &=& (i\omega_{c,L} - i\omega_R + \frac{\kappa_1 + \kappa_2}{2})\bar{a}_L + \\
    && ig_c(d_{01,L}\bar{\sigma}_{01,L} + d_{02,L}\bar{\sigma}_{02,L}) + it\bar{a}_R - \sqrt{\kappa_1}\bar{b}_{in,1}\\
    \Bar{\dot{a}}_{R} &=& (i\omega_{c,R} - i\omega_R)\bar{a}_R + \\ && ig_c(d_{01,R}\bar{\sigma}_{01,R} + d_{02,R}\bar{\sigma}_{02,R}) + it\bar{a}_L
\end{eqnarray*}

Analogously, the equations for the type-A setup of the 3-cavity system are:
\begin{eqnarray*}
    \Bar{\dot{a}}_{1} &=& (i\omega_{c,1} - i\omega_R + \frac{\kappa_1}{2})\bar{a}_1 + \\ && ig_c(d_{01,1}\bar{\sigma}_{01,1} + d_{02,1}\bar{\sigma}_{02,1}) + it_1\bar{a}_2 + it_3\bar{a}_3 - \sqrt{\kappa_1}\bar{b}_{in,1}\\
    \Bar{\dot{a}}_{2} &=& (i\omega_{c,2} - i\omega_R)\bar{a}_2 + \\ && ig_c(d_{01,2}\bar{\sigma}_{01,2} + d_{02,2}\bar{\sigma}_{02,2}) + it_1\bar{a}_1 + it_2\bar{a}_3\\
    \Bar{\dot{a}}_{3} &=& (i\omega_{c,1} - i\omega_R + \frac{\kappa_2}{2})\bar{a}_3 + \\ &&  ig_c(d_{01,3}\bar{\sigma}_{01,3} + d_{02,3}\bar{\sigma}_{02,3}) + it_3\bar{a}_1 + it_2\bar{a}_2
\end{eqnarray*}
While for the type-B setup of the 3-cavity system these expressions become:
\begin{eqnarray*}
    \Bar{\dot{a}}_{1} &=& (i\omega_{c,1} - i\omega_R + \frac{\kappa_1}{2} + \frac{\kappa_2}{2})\bar{a}_1 + \\ && ig_c(d_{01,1}\bar{\sigma}_{01,1} + d_{02,1}\bar{\sigma}_{02,1}) + it_1\bar{a}_2 + it_3\bar{a}_3 - \sqrt{\kappa_1}\bar{b}_{in,1}\\
    \Bar{\dot{a}}_{2} &=& (i\omega_{c,2} - i\omega_R)\bar{a}_2 + \\ && ig_c(d_{01,2}\bar{\sigma}_{01,2} + d_{02,2}\bar{\sigma}_{02,2}) + it_1\bar{a}_1 + it_2\bar{a}_3\\
    \Bar{\dot{a}}_{3} &=& (i\omega_{c,3} - i\omega_R)\bar{a}_3 + \\ && ig_c(d_{01,3}\bar{\sigma}_{01,3} + d_{02,3}\bar{\sigma}_{02,3}) + it_3\bar{a}_1 + it_2\bar{a}_2
\end{eqnarray*}

The set of equations above are complemented by those including the fields $\bar{b}_{out,i}$ as prescribed by the input-output theory.

\bibliographystyle{apsrev4-2}
\bibliography{refs}

\end{document}